\documentclass[preprint]{aastex6}
\usepackage{amsmath,amssymb,amstext}
\usepackage{graphicx, enumerate}

\defcitealias{2015ApJ...807....6C}{Paper I}

\shorttitle{Conduction effects in supra-arcade downflows}
\shortauthors{Zurbriggen et al.}

\begin{document}

\title{MHD simulations of coronal supra-arcade downflows \\ including anisotropic thermal conduction}

\author{E. Zurbriggen\altaffilmark{1,2}, A. Costa\altaffilmark{1,2,3}, A. Esquivel\altaffilmark{4}, M. Schneiter\altaffilmark{1,2,3}, M. C\'ecere\altaffilmark{1,2} }

\affil{$^1$ Instituto de Investigaciones en Astronom\'\i a Te\'orica y Experimental (IATE),  C\'ordoba, Argentina.\\
$^2$ Consejo Nacional de Investigaciones Cient\'\i ficas y T\'ecnicas (CONICET), Argentina.\\
$^3$ Facultad de Ciencias Exactas, F\'\i sicas y Naturales, Universidad Nacional de C\'ordoba (UNC), C\'ordoba, Argentina\\
$^4$ Instituto de Ciencias Nucleares, Universidad Nacional Aut\'onoma de M\'exico, Mexico}

\altaffiltext{1}{ezurbriggen@unc.edu.ar}



\begin{abstract}

Coronal supra-arcade downflows (SADs) are observed as dark trails descending towards hot turbulent fan shaped regions. Due to the large temperature values, and gradients in these fan regions the thermal conduction should be very efficient. While several models have been proposed to explain the triggering and the evolution of SADs, none of these scenarios address a systematic consideration of thermal conduction. Thus, we accomplish this task numerically simulating the evolution of SADs within this framework. That is, SADs are conceived as voided (subdense) cavities formed by non-linear waves triggered by downflowing bursty localized reconnection events in a perturbed hot fan. We generate a properly turbulent fan, obtained by a stirring force that permits control of the energy and vorticity input in the medium where SADs develop. We include anisotropic thermal conduction and consider plasma properties consistent with observations. Our aim is to study if it is possible to prevent SADs to vanish by thermal diffusion. We find that this will be the case, depending on the turbulence parameters. In particular, if the magnetic field lines are able to envelope the voided cavities, thermally isolating them from the hot environment. Velocity shear perturbations that are able to generate instabilities of the Kelvin-Helmholtz type help to produce magnetic islands, extending the life-time of SADs.
\end{abstract}


\section{Introduction} \label{s.intro}

The observational imprints of supra-arcade downflows (SADs)  are dark (subdense) moving trails descending from $\sim [40-60]~$Mm above the arcades, with downward speeds of $\sim [50 - 500]~$km s$^{-1}$ and sizes of $\sim [1 - 10]~$Mm \citep{1999ApJ...519L..93M,2012ApJ...747L..40S,2014ApJ...796...27I}. SADs have always been detected during long term erupting flares associated with CMEs \citep{2013ApJ...767..168L}. Furthermore, they have been observed  immersed in hot turbulent current sheet (CS) regions, with temperatures ($T \sim 10~$MK) much higher than the underlying coronal environment values \citep{2013ApJ...766...39M,2014ApJ...786...95H}, thus subject to strong gradients and large thermal conduction (TC) effects. Due to heat transfer, heat flowing from the high-temperature CS into the neighbouring plasma causes an expansion of the region called {\it thermal-halo} or {\it fan} (due to its shape).

In spite of the presence of large temperature gradients in SAD observations, TC effects are rarely considered explicitly in numerical simulations of fan regions where SADs develop \citep{2009EP&S...61..573L,2009MNRAS.400L..85C,2014ApJ...796L..29G}.  While several works on thermal effects in CS models have been developed (e.g., \citet{2008ApJ...689..572B, 2009ApJ...701..348S,2001ApJ...549.1160Y}), to the best of our knowledge, none of the numerical scenarios proposed to explain the SAD's origin and dynamics address the systematic inclusion of TC.

Within such framework, we proposed that SADs are voided cavities triggered by bursty reconnection events that occur during reconnection processes of longer term \citep{2009MNRAS.400L..85C} in inhomogeneous fans (\citealt{2012ApJ...759...79C}; and in \citealt{2015ApJ...807....6C}, hereafter \citetalias{2015ApJ...807....6C}). A bursty reconnection event deposits energy in a localized and instantaneous fashion, thus producing a blast wave. The blast generates shocks and rarefaction waves leading to the formation of the subdense cavity. These reconnection events were considered as local processes, i.e. independent of the overall fan region, that can be triggered outside the fan. For simplicity, the current simulations assume SADs are already immersed inside the fan. The bursty reconnection events are simulated as instantaneous pressure pulses. 

TC is a sensitive function of the temperature $(\propto T^{5/2})$ and is highly anisotropic in the presence of magnetic fields \citep{1962pfig.book.....S}, where the heat flux is mostly funnelled along the field direction. Therefore, in an environment with magnetic fields and large thermal gradients, TC is highly efficient along the magnetic field lines, but almost null in the perpendicular direction. For this reason, almost closed magnetic field structures are left as the only efficient mechanism to hinder TC. In this setting, \citet{2007A&A...464..753P} studied the evolution of CME cores embedded in a hot coronal plasma using the magnetic field as thermal insulation to suppress TC.

In the context of  the SAD and their environment \citetalias{2015ApJ...807....6C} showed that TC can only be neglected if the fan densities and the characteristic longitudes are large enough. This assumption implies a fan where  $\beta > 1$ (the ratio of thermal pressure to magnetic pressure). Also, \citet{2013ApJ...766...39M} and \citet{2016ApJ...819...56S} found relatively large values of $\beta$ in the corona
($> 1$) suggesting that either the density and/or the temperature are larger than generally assumed, or that the magnetic field is small enough. Thus, the fine determination of the plasma parameters in SADs environment is a key question when trying to solve the  importance of TC in this context, as well as the configuration of the magnetic field, since it might behave as a thermal insulator, allowing SADs to last longer (e.g. see figure 4 of \citet{2004ApJ...605L..77A}).
   
Thus, as TC modifies the flow dynamics when its time-scale is comparable to (or smaller than) the Alfv\'en time-scale, 
models based on the development of instabilities, as proposed by \citet{2014ApJ...796L..29G}, require a thorough discussion of TC effects. In this sense, \citet{2002Ap&SS.281..275H} studied the suppression of hydrodynamic instability due to TC.
   
In what follows, the scenario proposed in \citetalias{2015ApJ...807....6C} is revised, taking into account the effects of anisotropic TC in a turbulent magnetized fan, i.e.,  the environment where SADs are observed. To summarize, the aim of this study is to address the following questions: how is it possible that SADs resist the thermal diffusion and are observed in hot fan regions? In fact, perturbations associated with typical SADs, --of sizes of various Mm, typical coronal ion number densities of $\sim 10^{9}~$cm$^{-3}$ and temperatures as high as $\sim [7-13]~$MK-- will fade away in times (of a few seconds) that are at least two orders of magnitude smaller than a typical SAD life-time. Is there a mechanism to avoid the thermal diffusion?
The model is presented in  Section 2, followed by a description of the simulations in Section 3. The results can be found in Section 4, and in section 5 we provide our conclusions.

\section{The Model} \label{s.model}

We consider MHD equations in conservative form (CGS units): 
\begin{eqnarray}
\frac{\partial \rho}{\partial t} + \nabla \cdot \left(\rho\mathbf{v}\right)  &=& 0, \label{e.density} \\ 
\frac{\partial \rho  \mathbf{v}}{\partial t} + \nabla \cdot \left(\rho\mathbf{v}\mathbf{v}-\mathbf{B}\mathbf{B}\right) + \nabla p_* &=& \mathbf{f_{force}}, \label{e.momentum} \\ 
\frac{\partial E}{\partial t} + \nabla\cdot \left[(E+p_*) \mathbf{v}-\mathbf{B}(\mathbf{v}.\mathbf{B})+\mathbf{F_{c}}\right] &=& 0,  \label{e.energy} \\ 
\frac{\partial \mathbf{B}}{\partial t} = \nabla\times \left(\mathbf{v}\times\mathbf{B}\right), &&  \label{e.induction} \\ 
E = \frac{1}{2}\rho v^2 + e + \frac{B^2}{2},  &&  \\ 
p_* = p+\frac{B^2}{2}, && \\
p = (\gamma -1)e = \frac{R}{\bar{\mu}}\rho T, &&
\end{eqnarray}
where $\rho$ is the mass density, $t$ is the time, $\mathbf{v}$ is the plasma flow velocity, $p$ is the thermal pressure, $\mathbf{B}$ is the magnetic field divided by $\sqrt{4\pi}$, $E$ is the total energy, $e$ is the internal (thermal) energy, $\gamma=5/3$ is the ratio of the specific heats ($C_P/C_V$), $R$ is the gas constant, $\mathbf{F_c}$ is the heat flux,
and $\mathbf{f_{force}}$ is a turbulence driving force. The plasma is assumed to be completely ionized with solar abundance\footnote{Solar abundance: $70.7\%$ H $+ 27.4\%$ He $+ 1.9 \%$ heavier elements \citep{2000itss.book.....P}.}. Hence, the density is $\rho=(n_{i}+n_{e})m_{H}\bar{\mu}$, where $n_{i}$ is the ion numerical density, $n_{e}=\bar{z}n_{i}$ is the free electron numerical density, $\bar{\mu}=0.613$ is the mean atomic mass, $\bar{z}=1.445$ is the mean atomic number and $m_{H}$ is the hydrogen atomic mass.
\begin{table}
\begin{center}
\begin{tabular}{@{}ccccccccc}
\hline  \hline
       {}           & $T$ [MK] &  $\rho$ [gr cm$^{-3}$]    & $\mathbf{B}\cdot\mathbf{\hat{j}}$ [G]  & $|\mathbf{v}| \ [c_{s}]$  &  Stir & TC & Shear \\ \hline  
\textbf{Turb} &   $6.0$   & $1.2\times 10^{-14}$      &                             $3.0$                            &               $0.0$                 &   yes   &   no  &  no   \\ 
\hline  
\textbf{Fan}   & $\langle7.0\rangle$  & $\langle1.2\times 10^{-14}\rangle$ &                         $\langle3.0\rangle$                   &         $\langle0.21\rangle$   &  no   & yes & yes  \\ \hline
\end{tabular}
\caption{Second row: homogeneous initial conditions of the simulation and of the turbulent stage. Third row: initial conditions of the fan stage, i.e. the plasma parameters once the turbulent stage is completed (framework parameters for the 2D simulations). The initial density $\rho$ is equivalent to $n_{i}=5\times 10^{9}~$cm$^{-3}$. The quantities in $\langle\dots\rangle$ means average over all grid cells. At the initial fan stage $<c_{s}>=4.0\times 10^{7}$cm~s$^{-1}$.}
\label{table.uno}
\end{center}
\end{table}

As here we are interested in the relation between SADs and TC in the framework of the scenario proposed in previous papers  (\citealt{2009MNRAS.400L..85C, 2012ApJ...759...79C}; and \citetalias{2015ApJ...807....6C}), to perform the runs we assume a set of characteristic parameters which lead to subdense cavities affected by TC when a pressure pulse in a 2D background turbulent fan is imposed. The set of values is given in  Table (\ref{table.uno}). Also, \citet{2012ApJ...747L..40S} showed that for a fan with temperatures in the range of  $ \sim [10 - 13]~$MK  the SAD emission measure (EM) contrast has to be at least a factor of two with respect to the surroundings (see figure 4 of their paper). 

Based on these observational constraints, and in order to numerically reproduced a detectable SAD, we know that the following conditions have to be satisfied: 1) an EM contrast between the subdense cavity and its background of at least a factor of $4$; and 2) this subdense cavity must last at least one minute in the turbulent background.

To calculate the EM we make use of the approximation of optically thin plasmas,
\begin{equation} \label{e.EM}
EM=\int_{l_i}^{l_e} n_{e}^{2} \ dl.
\end{equation}
The EM quantifies the squared electronic density $n_e$ along the line of sight direction $\vec{l}$ of the observer. In what follows, the line of sight is assumed to be along the $z$-direction. As we perform $2D$ simulations in the $(x,y)$ plane, the EM evaluation is $EM/(z_e-z_i)=n_e^2$ where we assume that the spatial dependence of the number density is $n_e(x,y)$. In that sense the EM that we present can be considered as the emission measure per unit length along the line of sight.

Some useful characteristic time-scales can be calculated in order to estimate the relative importance of various physical processes in a typical fan. The conduction, radiative and magnetic reconnection time-scales are respectively:
\begin{eqnarray} \label{e.time-scales}
t_{cond} &=&9.1\times 10^{13}\frac{\rho L^2}{T^{2.5}}, \quad  t_{rad} = \frac{3}{2}\frac{P}{\bar{z}n_i^2 E_{r}}, \\
t_{rec} &=& \frac{L}{v_{A} M_{A}^2}, \nonumber
\end{eqnarray}
where $L$ is a characteristic length and $E_{r}$ is the radiative loss function ($E_{r}\simeq 4\times 10^{-23}~$erg cm$^{3}$ s$^{-1}$ for $T\backsimeq 10^7~$K \citep{2005psci.book.....A}). For the reconnection time-scale we assume that the reconnection is mediated by turbulence, following
\citet{1999ApJ...517..700L} where $v_{A}$ is the Alfv\'en speed and $M_{A}$ is the Mach Alfv\'en number. When considering the set of plasma parameters of the current problem the corresponding time-scales are: $t_{cond}\sim 0.5~$s, $t_{rad}\sim 10^3~$s, $t_{rec}\sim 10^3~$s.  Note that, excepting for $t_{cond}$, the time-scales are of the same order of magnitude as a typical SAD life-time, i.e. only the TC is capable of rapid modification of the SAD observability. SADs  maintain a pressure balance with their neighbourhood because they are hotter than the surrounding plasma, hence, unless a mechanism acts to  inhibit the strong TC diffusion, SADs will not last for times comparable with the observations.

Estimations of the Kelvin-Helmholtz (KH) and Rayleigh-Taylor (RT) instabilities are given by
\begin{equation} \label{e.khrt_g}
t_{KH-RT}\equiv \frac{\gamma}{\gamma-1}\frac{c_{s}}{g_{Sun}},
\end{equation}
or
\begin{equation} \label{e.khrt_p}
\tau_{KH-RT}\equiv  \frac{\gamma}{\gamma-1}\frac{d \ \ c_{s}}{v^2},
\end{equation}
being $c_{s}$ the sound velocity, $g_{Sun}$ the solar gravitational acceleration at the low corona, $d$ and $v$ are typical SAD widths and speeds. 
Eq. (\ref{e.khrt_g}) gives the time-scale taking into account the Sun gravity \citep{1972SoPh...25..380C}. Within the set of our framework parameters $t_{KH-RT}\sim 3\times 10^3~$s, thus the conduction terms will diffuse the structures formed by these type of instabilities, making them unlikely detectable as SADs. On the other hand, \citet{2014ApJ...796L..29G} generated SADs by means of Rayleigh-Taylor type instabilities produced by reconnection downflows that exert a ram pressure between an above lighter plasma and a below denser fan region. In this case, Eq. (\ref{e.khrt_p}) gives the time-scale taking into account an acceleration $v^2/d$ that plays the role of gravity. \citet{2014ApJ...796L..29G} obtained SADs of widths $d \sim [5-12]~$Mm and average speeds of $50~$km s$^{-1}$ leading to a $\tau_{KH-RT}\sim  10^3~$s.
Therefore, the (isotropic) TC would also produce a strong diffusion in structures formed by these type of instabilities. However, $\tau_{KH-RT}$ could be reduced considering larger and also typical SAD speeds $v$; additionally SADs could be thermally isolated by their particular magnetic field configuration.
In \citetalias{2015ApJ...807....6C} we argued that the source and sink terms (i.e., TC, radiation losses and reconnection contribution of heat)  could be ignored if  they compensate each other. To show that this could be the case  we considered somewhat large values of the number density, $n_i=2 \times 10^{10}\,$cm$^{-3}$, and of the characteristic longitude $L=12\,$Mm. However, observational data also report SADs sizes of $\sim1\,$Mm that remain visible for several minutes, which cannot be reproduced within our model including TC, because of their $t_{cond}$ are even shorter (see the discussion of Fig. (\ref{f.EM_contrast})). 	

The bursty reconnection events are simulated as instantaneous pressure pulses. They are generated as sudden thermal pressure increases, not modifying the density at the pulse region. Consequently the SAD temperature results larger than its neighbourhood values, allowing an almost total pressure equilibrium between the SAD structure and its near neighbourhood. Our results from two-dimensional simulation seem to disagree with \citet{2009ApJ...697.1569M}, who suggested that the SAD collapse is avoided due to higher internal values of the magnetic pressure, but we need three-dimensional simulation to confirm this conclusion.
  Due to the simplified SAD setup proposed here (where the medium outside the fan is not modelled) we can only consider the situation where SADs are triggered already inside the fan. \citet{2014ApJ...786...95H} pointed out that there is little evidence that SADs are hotter than the fan, but they are always hotter than the external background medium. 
Definitely a whole magnetic configuration (considering the fan and the background medium) with SADs generated outside and descending into the fan, might result in SADs with internal temperatures higher than the background medium but colder than the fan temperature values.    

The purpose of the present work is to investigate whether different combination of the parameters $(\rho,T,L)$, i.e. lower values of the density and smaller SAD sizes, together with additional physical mechanisms that prevent the action of the heat diffusion terms could explain the observed SAD characteristics. That is, considering the high efficiency of the TC in the fan, we explore situations where the magnetic field lines could envelope a SAD, providing thermal insulation from its surroundings.

\section{Numerical code and initial conditions} \label{s.ic}

To carry out the numerical simulations we use the {\sc flash} code (\citealt{2000ApJS..131..273F}, release 4.2.2) that solves the compressible MHD equations. We choose for our simulations the unsplit staggered mesh algorithm \citep{2009ASPC..406..243L} available in {\sc flash}, which employs a finite volume method with a directionally unsplit data reconstruction and the constraint transport method to enforce the $\nabla \cdot \mathbf{B} = 0$ condition. The Riemann problems at the computational interfaces are calculated using the second order Roe Riemann-solver along with a  MC slope-limiter. A uniform Cartesian $2D$ grid with $300^2$ cells is used. The physical domain representing the fan is set up to $(15,15)~$Mm, with the $y$-coordinate pointing away (radially from the Sun, neglecting the curvature of its surface) and the $x$-coordinate parallel to the surface of the Sun. Periodic conditions are considered at all boundaries.

The preparation of the fan setup where the SADs are triggered is performed in two stages. Starting from a homogeneous  initial conditions given in the second row of Table (\ref{table.uno}) we generate a turbulent state. This stage is obtained by adding a driving force that generates a turbulent state as is  described in the next section. At the end of the stirring stage the system reaches a Kolmogorov-like spectrum with the plasma parameters showed in the third row of Table (\ref{table.uno}). This turbulent state is the initial condition of the second (fan) stage. During the fan stage we turn off the stirring force leaving the turbulence to gradually decay, but we also turn on the anisotropic TC, as well as a shear resembling the action of outflows coming from above the fan. After this stage we impose four bursty reconnection pulses. The turbulent stage occupies the time interval $t_{turb}=[0-300]\,$s while the fan stage takes $100$ additional seconds ($t_{fan}=[300-400]\,$s). The shear perturbation is instantaneously applied in the $y$-component of the velocity in the form
\begin{equation}
v_{y}(t,x,y) = \left\{
\begin{array}{l l}
 v_{y}(t,x,y) - v_{sh}  & \quad \mbox{if $ x < 0 $,} \\
 v_{y}(t,x,y)          & \quad \mbox{if $ x \ge 0$, }
\end{array} \right.  \label{10}
\end{equation}
at time $t=303.8~$s and with $v_{sh}=1.5\times 10^{7}\,$cm s$^{-1}$. Finally the four bursty reconnection pulses are instantaneously applied at time $t=305.9~$s, with diameters $L=1.2\,$Mm, labelled as $(A)$, $(B)$, $(C)$, $(D)$, centred at $(-3.0\times 10^{8},-4.9\times 10^{8})\,$cm, $(-9.6\times 10^{6},-3.1\times 10^{8})\,$cm, $(5.3\times 10^{7},5.5\times 10^{8})\,$cm, and $(5.1\times 10^{8},4.3\times 10^{8})\,$cm, and with pressure contrasts of $\Delta P/P=(5,4,5,4)$, respectively.

In order to include the anisotropic TC (see for instance \citealt{1962pfig.book.....S}), a new module in the {\sc flash} code was developed. In this new module the advected energy flux and the heat flux $\mathbf{F_c}$  were combined into a new energy flux inside the divergence at the left hand side of Eq. (\ref{e.energy}). The components of the anisotropic thermal flux $\mathbf{F_c}$ depend on the relative direction between the magnetic field and the temperature gradient as follows:
\begin{eqnarray}
\mathbf{F_c} &=& - q_{eq_{\parallel}}\nabla T_{\parallel} - q_{eq_{\perp}}\nabla T_{\perp}, \\
\nabla T_{\parallel} & = & (\mathbf{\hat{b}}.\nabla T)\mathbf{\hat{b}}, \qquad  \nabla T_{\perp} = (\nabla T - \mathbf{\hat{b}}.\nabla T)\mathbf{\hat{b}}, \\
q_{eq_{\parallel}} &=& \left(\frac{1}{q_{spi_{\parallel}}}+\frac{1}{q_{sat_{\parallel}}}\right)^{-1}, \\
q_{eq_{\perp}} &=& \left(\frac{1}{q_{spi_{\perp}}}+\frac{1}{q_{sat_{\perp}}}\right)^{-1}, \\
q_{spi_{\parallel}} &=& 6.4\times 10^{-7}T^{2.5}, \qquad q_{sat_{\parallel}} = \frac{5\phi\rho c_s^3}{|\nabla T_{\parallel}|}, \\
q_{spi_{\perp}} &=& 4.6\times 10^{32}\frac{\rho^2 }{B^2 \sqrt{T}}, \quad q_{sat_{\perp}} = \frac{5\phi\rho c_s^3}{|\nabla T_{\perp}|},
\end{eqnarray}
where the subscripts $\parallel$ and $\perp$ refer to the directions along and across the magnetic field, respectively. $\nabla  T$ is the temperature gradient, $\phi$ is a parameter that we set to $\phi=0.3$, ${q_{spi_{\parallel}}}$ and ${q_{spi_{\perp}}}$ are the classical thermal conductivity \citep{1962pfig.book.....S}, ${q_{sat_{\parallel}}}$ and ${q_{sat_{\perp}}}$ are the saturated thermal conductivity \citep{1977ApJ...211..135C}, $q_{eq_{\parallel}}$ and $q_{eq_{\perp}}$ are the equivalent thermal conductivity resultant of applying an harmonic flux-limit. 

Thermal conduction imposes a strong diffusive constraint to the time-steps. The constraint is a parabolic-like one, $\Delta t_{diff} < 0.5\rho C_{V}\Delta x^2/max(q_{eq_{\parallel}},q_{eq_{\perp}})$, where $C_{V}=R/\left( (\gamma-1)\bar{u}\right)$ is the specific heat at constant volume and $\Delta x$ is the cell size. Therefore, during the simulation runtime each time-step value is determined by the minimum between the usual CFL constrain and the diffusive constraint, i.e. $\Delta t=\min(\Delta t_{CFL},\Delta t_{diff})$. The relatively low temperature value chose for the initial fan stage ($T=7~$MK) is a compromise between typical observational values and the computational demanding diffuse constraint, $\Delta t_{diff}$. 

Although for all cases explored (i.e. the parameter ranges $\rho \sim [10^{-14}-10^{-15}]~$gr cm$^{-3}$, $T \sim [1-10]~$MK, $|\mathbf{B}| \sim [3-6]~$G) the equivalent thermal conductivities always satisfied the condition $q_{eq_{\parallel}} \gg q_{eq_{\perp}}$, both contributions were taken into account.

\subsection{The role of thermal conduction} \label{sub.tc}

\begin{figure} 
\begin{center}
\includegraphics[width=0.4\textwidth]{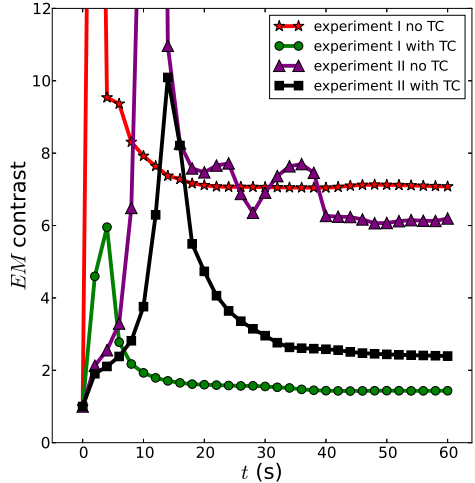}
\caption{EM contrast against time for the blast experiments I and II  with and without anisotropic TC.}
\label{f.EM_contrast}
\end{center}
\end{figure}

To illustrate the TC effects over a subdense cavity we carry out a simple $2D$ MHD blast test, neglecting and including anisotropic TC. We use a $200^2$ cells setup. The initial density and pressure are homogeneous excepting in a central circle of diameter $L$, where the pressure is increase by $\Delta P$. The fluid is initially at rest. A uniform magnetic field is assumed along the y-direction. In Fig. (\ref{f.EM_contrast}), we show the results of two numerical experiments:
\begin{enumerate}[I)]
\item Our present framework parameters, based on observations of long-lived small SADs: 
 $\rho=1.2\times 10^{-14}\,$gr~cm$^{-3}$, $T=7\,$MK, $\mathbf{B}\cdot\mathbf{\hat{j}}=3\,$G, $\mathbf{v}=0$, $L=1.2\,$Mm and $\Delta P/P=5$
\item With the parameter range explored in \citetalias{2015ApJ...807....6C}: $\rho=5\times 10^{-14}\,$gr cm$^{-3}$, $T=10\,$MK, $\mathbf{B}\cdot\mathbf{\hat{j}}=5\,$G, $\mathbf{v}=0$, $L=12\,$Mm and $\Delta P/P=4\,$.
\end{enumerate}
Both setups lead to a blast that generates a subdense cavity formed by shocks and expansion waves, which is faded away by the anisotropic TC. We define the EM contrast function, at a given time, as the maximum ratio between the background EM (almost equal to the unperturbed EM value) and its value inside the voided cavity, i.e. the EM $\mathrm{contrast}(t)=\max\left[EM_{back}/EM_{cavity}(t)\right]$ whose evolution can be analysed in view of the TC time-scale. This function gives a rough estimation of the permanence of the EM contrast along the time. Figure (\ref{f.EM_contrast}) shows the EM contrast as a function of time for both numerical experiments (I and II) with and without TC. Note that for large times the EM $\mathrm{contrast}(t) > 4$ for both cases that do not include TC (red and purple lines), meaning that a SAD is detectable according to the constrains given above.  However, SADs will not be detected when TC is considered (green and black lines). In spite of this, the case equivalent to \citetalias{2015ApJ...807....6C}, with large values of $L$ and $\rho$, is in the limit of detection possibilities.
%

\subsection{Forced turbulence} \label{sub.turbulence}

Motivated by observational reports of the turbulent flows and vortices in fans \citep{2013ApJ...766...39M, 2016ApJ...819...56S}, as well as the SADs detections in highly perturbed fan regions (see \citealt{2012ApJ...747L..40S} movies), we may expect that a turbulent fan is the natural site where SADs develop.

To generate a statistically stationary turbulent state we force the low wave-number (i.e. large scale) velocity components by artificially injecting energy. The energy cascades down systematically to smaller scales until it is finally  dissipated by viscous action at the smallest scales. Starting from an initial condition, a statistically steady state is achieved after some time in which the average rate of energy added to the system balances with the average energy-dissipation rate \citep{1988CF.....16..257E}. To force the system we use a numerical device that emulates a stirring process.

In order to generate a proper turbulent background we make use of the {\it Stir} unit ({\it FromFile} implementation) in the {\sc flash} code \citep{2010A&A...512A..81F}. The {\it Stir} unit allows to drive a solenoidal force that is calculated in Fourier space and then added as a source term to the momentum equation. This force is represented by $\mathbf{f_{force}}$ in the right hand side of Eq. (\ref{e.momentum}). 

The turbulence is driven in a range of wave-numbers that have a parabolic distribution that covers scales from roughly one third of the computational box ($k=3$) to the whole computational box ($k=1$), peaking at $k_{peak}=2$.
The driving amplitude  ($e_{inj}=2.0\times 10^{10}~$(erg/s)$^{1/2}$), was chosen so that the mean temperature after the turbulence reaches a quasi-steady state does not exceed $7\,$MK. The auto-correlation time-scale $t_{corr}$ of the forcing, which is equal to the dynamical time-scale of the vortices excited, $t_{corr}=L_{peak}/v = 40.1~$s, with $L_{peak}=2\pi/k_{peak}$, and an arbitrarily chosen $v=c_{s}/2$. We carry out the forcing in a pure solenoidal mode (divergence-free) and for the time between successive driving patterns the recommended $10\%$ of the correlation time is used, i.e. the number of  driving patterns $n_{patt}$ used during runtime is $n_{patt}=10 t_{turb}/t_{corr}$, where $t_{turb}=7.5 t_{corr}=300\,$s is the turbulent stage duration. 
\begin{figure}
\begin{center}
\includegraphics[width=0.4\textwidth]{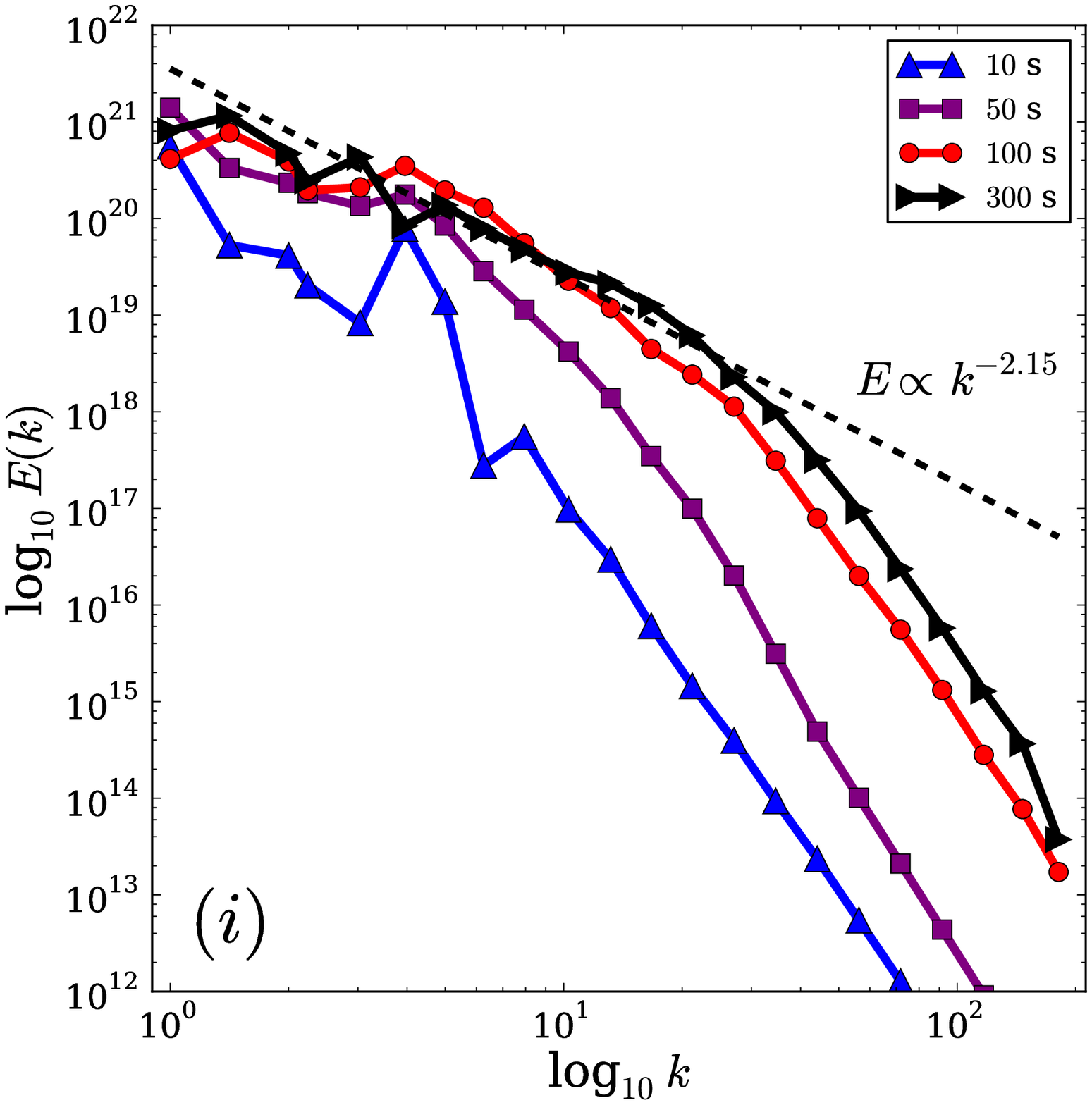} 
\includegraphics[width=0.4\textwidth]{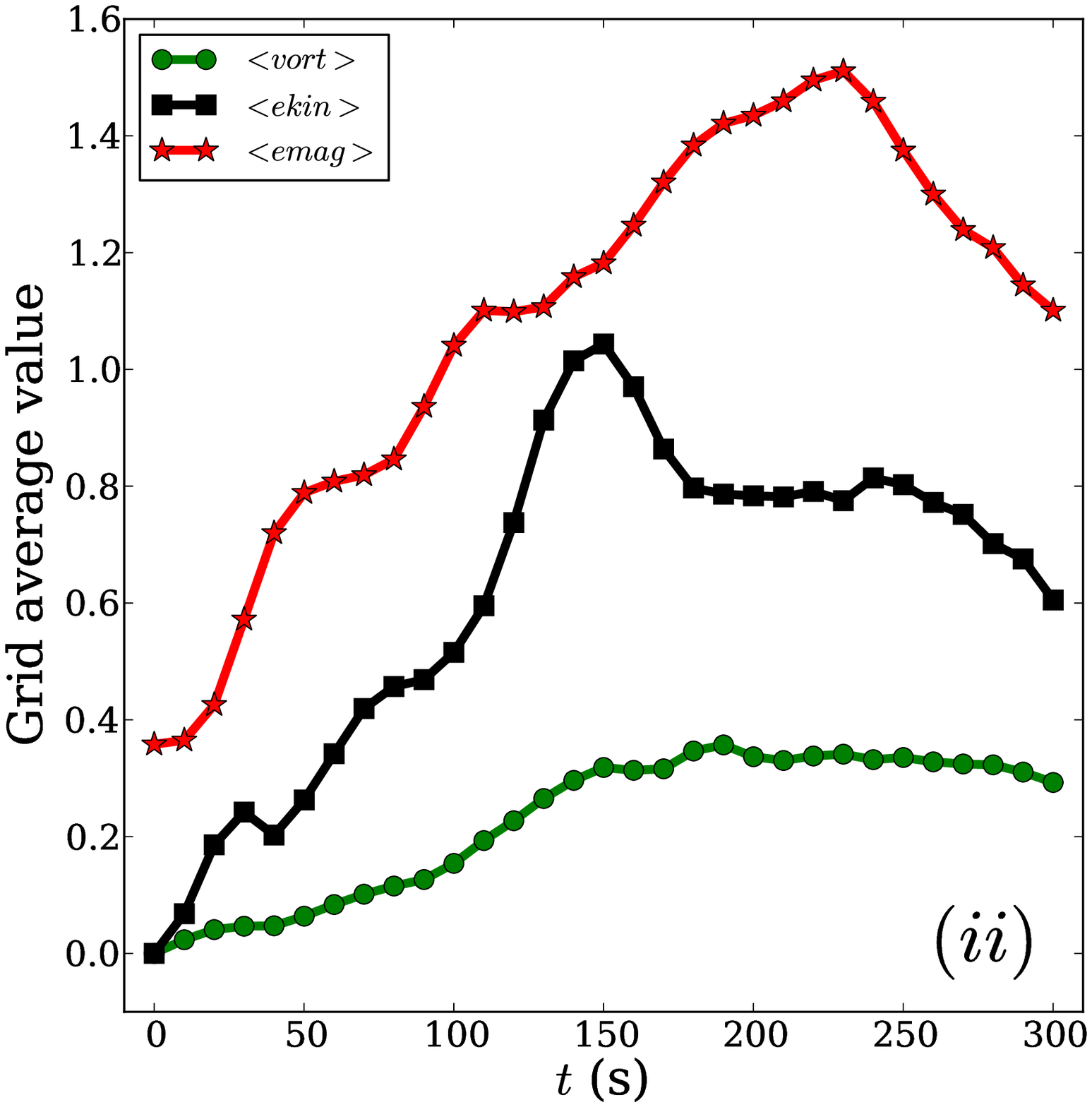}
\caption{(i) Velocity spectral evolution, $E(k)$, for different times during the turbulent stage. (ii) Vorticity, magnetic energy and kinetic energy evolution. (Parallel magnetic field case).}
\label{f.turb}
\end{center}
\end{figure}

In a nutshell, starting from a homogeneous rest a stationary turbulent state is generated, considered as the initial condition for the fan medium. Figure \ref{f.turb}(i) shows the Fourier power spectrum of the velocity field $E(k)$ for different evolution times during the stirring stage. Figure \ref{f.turb}(ii) shows the grid cell average value of the vorticity, kinetic energy and magnetic energy against time for stirring runtime. At early times in this stage the kinetic energy and vorticity have low average values while the velocity power spectrum revels that large scales (low wave-number) are excited but small scales (high wave-number) are not yet perturbed (blue curve in Fig. \ref{f.turb}(i)). That is, a stationary state has not been reached. Later on the energy cascade occupies its entire inertial range and the system reaches a stationary state. After this the velocity power spectrum and the other average quantities become stable, with only small oscillations around mean values. Since no additional energy sinks (other than numerical dissipation) were considered and periodic boundary conditions were used, at the beginning the total energy increases monotonically (almost linear) with time. In the stationary state, we find the power spectrum of the velocity scale as $E(k)\propto k^{-2.15}$ within the inertial range, which is in rough agreement with the expected  scaling of a $2D$  Kolmogorov-like turbulence $(E(k) \propto k^{-8/3})$\footnote{Kolmogorov turbulence corresponds to turbulence in incompressible unmagnetized medium, which is not the case of the solar corona. The fact that we have such spectrum stems from the fact that shear Alfv\'en modes (which are incompressible) dominate the cascade \citep{1995ApJ...438..763G}.}. We take as the numerical inertial range $k_{5}\lesssim k\lesssim k_{15}$, adopting the criterion followed by \citet{2010A&A...512A..81F} (and references therein).

When the stationary state is achieved the kinetic, the vorticity and the magnetic functions cease to grow. This stationary state is generated for two initially different magnetic fields configurations: a homogeneously parallel (Fig. \ref{f.turb}(ii)) and an anti-parallel (with respect to the radial direction from the Sun). The anti-parallel case defines the position of the CS. As the CS is expected to be a very thin region ($\sim 1\,$Mm) surrounded by the fan itself and vortical features were apparently detected all over the fan region \citep{2013ApJ...766...39M, 2016ApJ...819...56S}, we assume that eddies must be formed independently of the location of the CS. Thus, we perform the analysis using the more general parallel configuration and we will come back later to the discussion of the anti-parallel configuration.

\section{Results and Discussion} \label{s.results}

\begin{figure} 
\begin{center}
\includegraphics[width=0.4\textwidth]{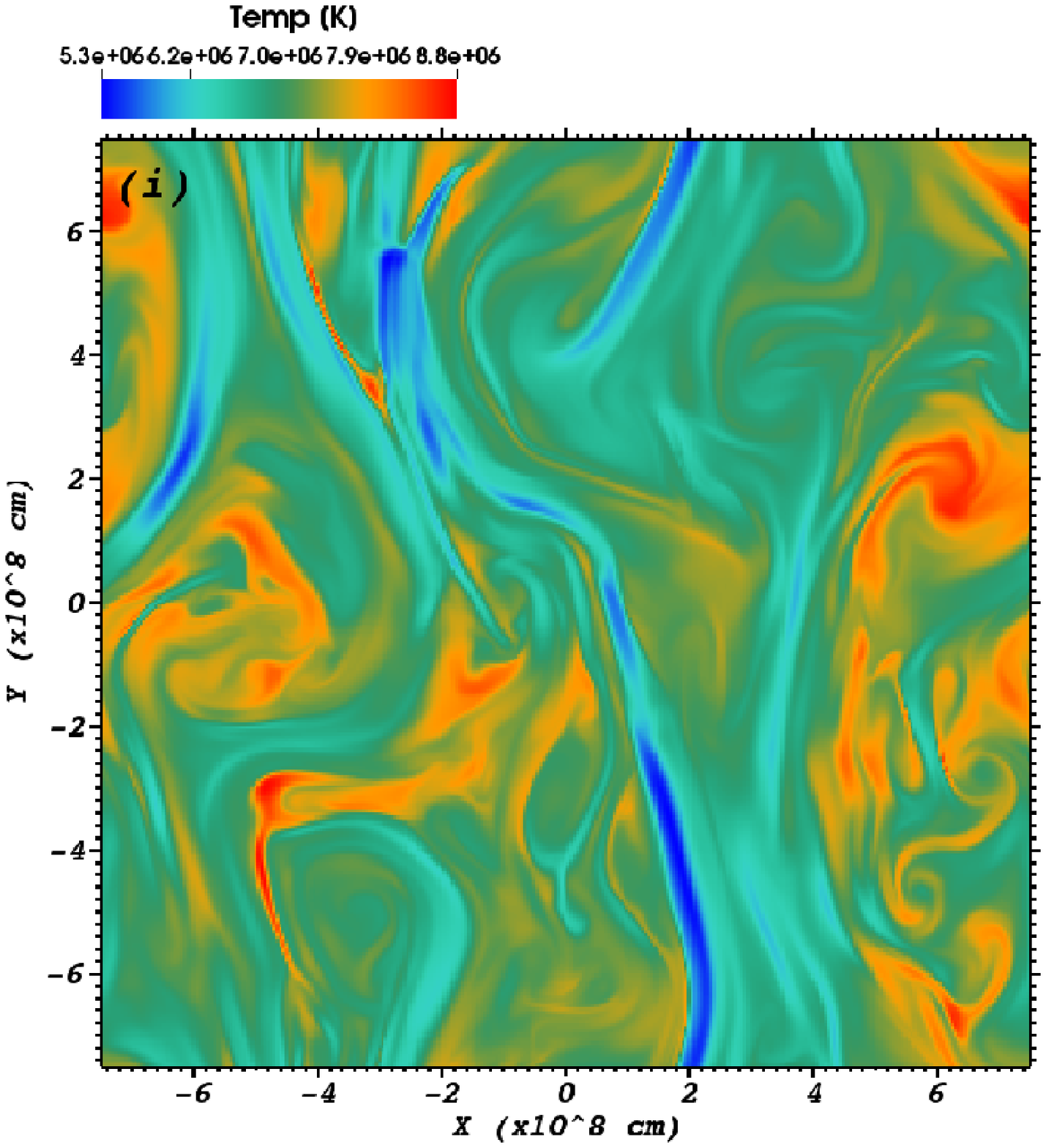}
\includegraphics[width=0.4\textwidth]{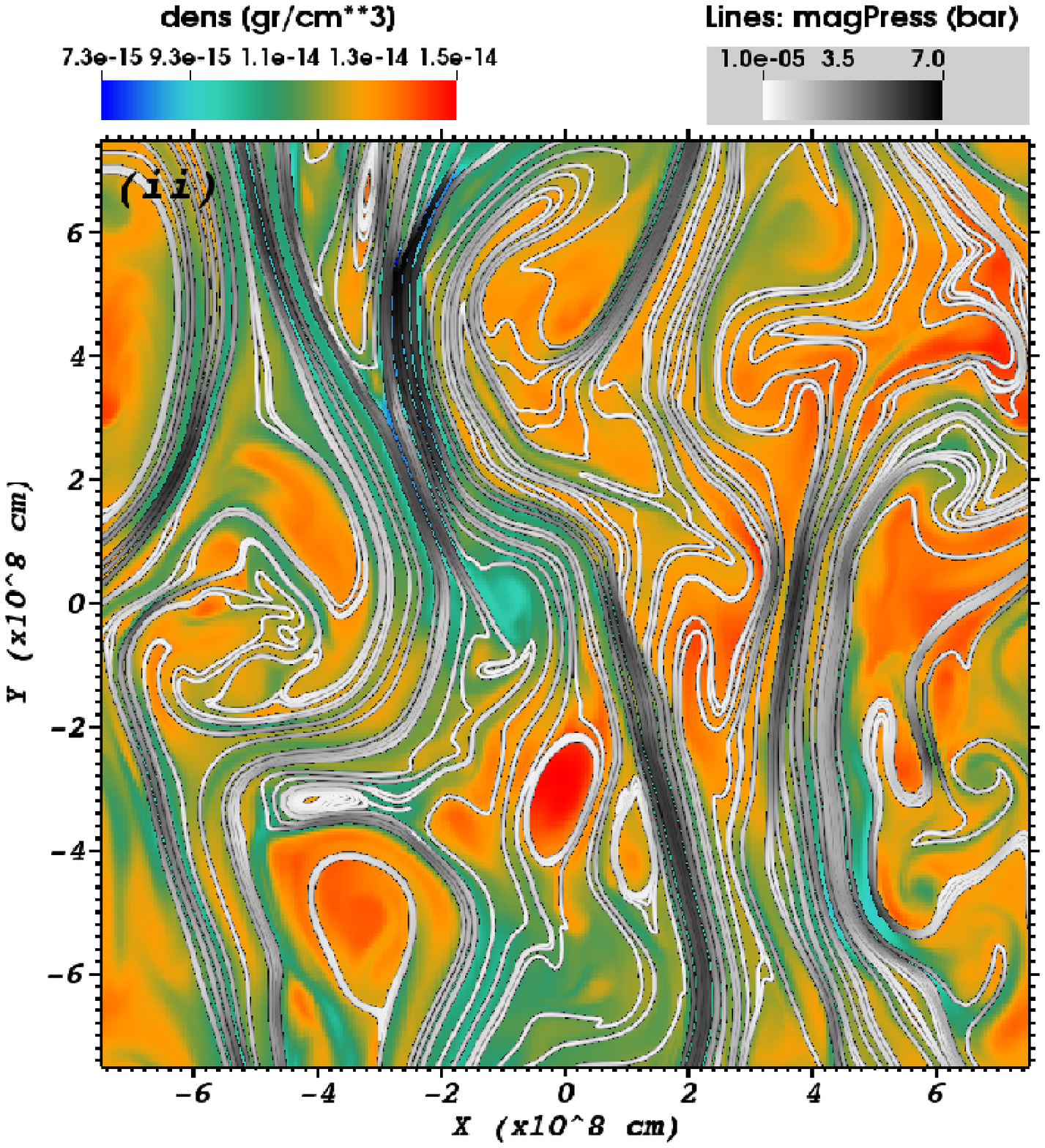}
\caption{(i) The temperature once the turbulent stage ends, at $t=300~$s. (ii) Same as (i) but for the density superimposed with magnetic field lines.  This is the initial condition for the fan stage.} 
\label{f.ic}
\end{center}
\end{figure}

Once the stationary turbulent state is established we turn off the forcing mechanism, corresponding to the initial conditions of the fan stage. Figure \ref{f.ic}(i) displays the density with overlapped magnetic field lines, and Fig. \ref{f.ic}(ii) shows the temperature for the stationary state which is obtained ensuring a final temperature of $\sim 7\,$MK, the desired SAD framework temperature.

At this time, the magnetic topology forms different structures, strands and vortices. Strands are accumulations of magnetic field lines generally associated with relatively low internal values of temperature and density, meaning that the low gas pressure is compensated by large values of the magnetic pressure \citep{2013ApJ...766...39M}. Vortices tend to possess closed magnetic topologies of approximately homogeneous internal density and temperature. The high density regions, partitioned by the magnetic field lines, correspond either to relatively high  or to relatively low temperature. Also {\it islands} (regions with totally closed magnetic field line) are formed. As in \citet{2013ApJ...766...39M,2016ApJ...819...56S}, the fan has an average initial state value $\beta\gtrsim 1 \sim 28$. With this value of $\beta$ the turbulence is almost hydrodynamic, as the magnetic forces do not play a dominant role in the dynamics.

Then we turn on the anisotropic TC. As the vortical regions are the candidates to thermally isolate possible SADs we apply pressure pulses close to vortices. For simplicity we are not simulating the entrance of SADs to the fan. We assume that as SADs propagate into the fan and their motions are such that they become wrapped up by the magnetic field lines or that, eventually, local reconnections occur inside the fan when two non-parallel field lines approach enough. 

\begin{figure} 
\begin{center}
\includegraphics[width=0.4\textwidth]{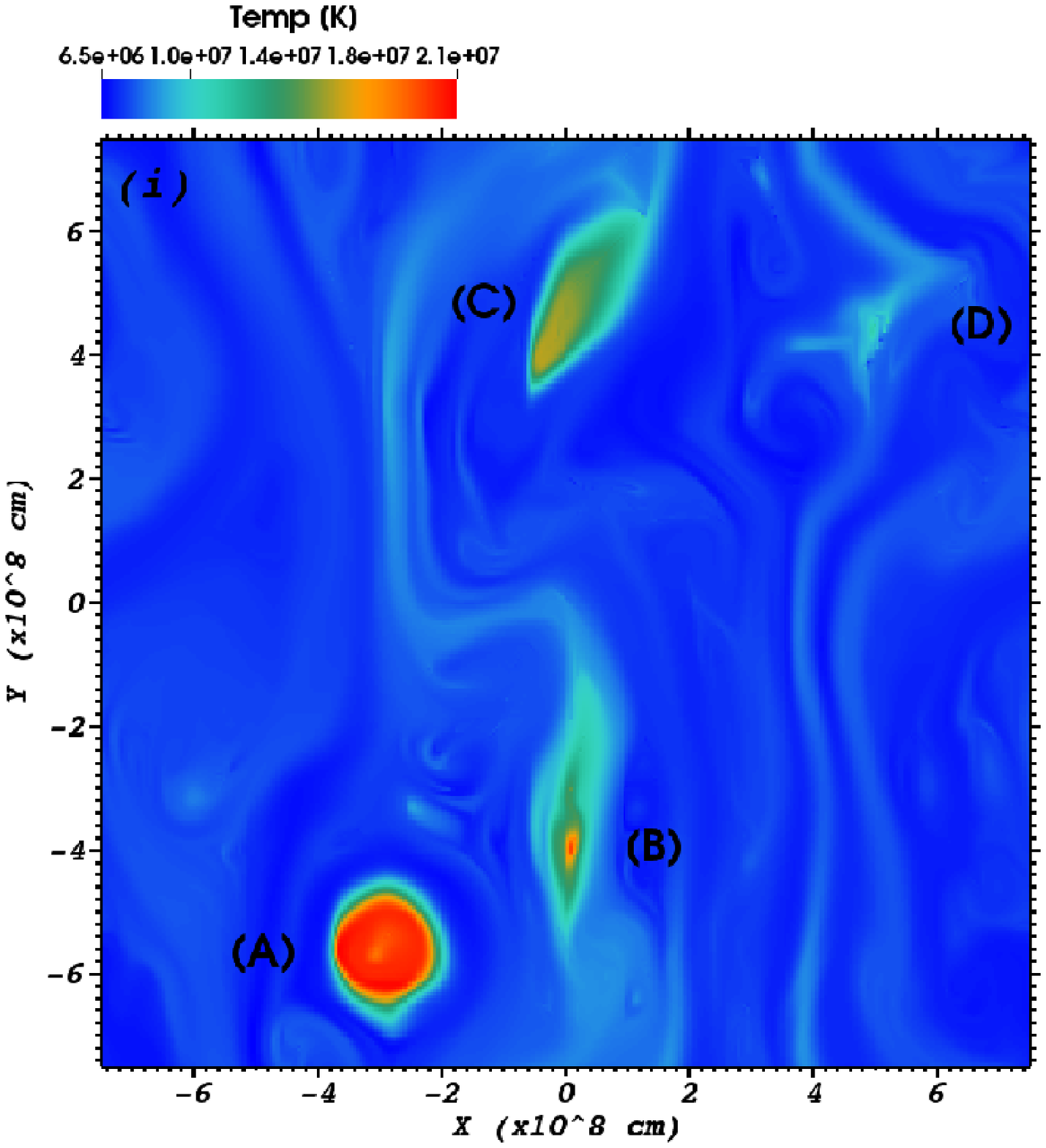}
\includegraphics[width=0.4\textwidth]{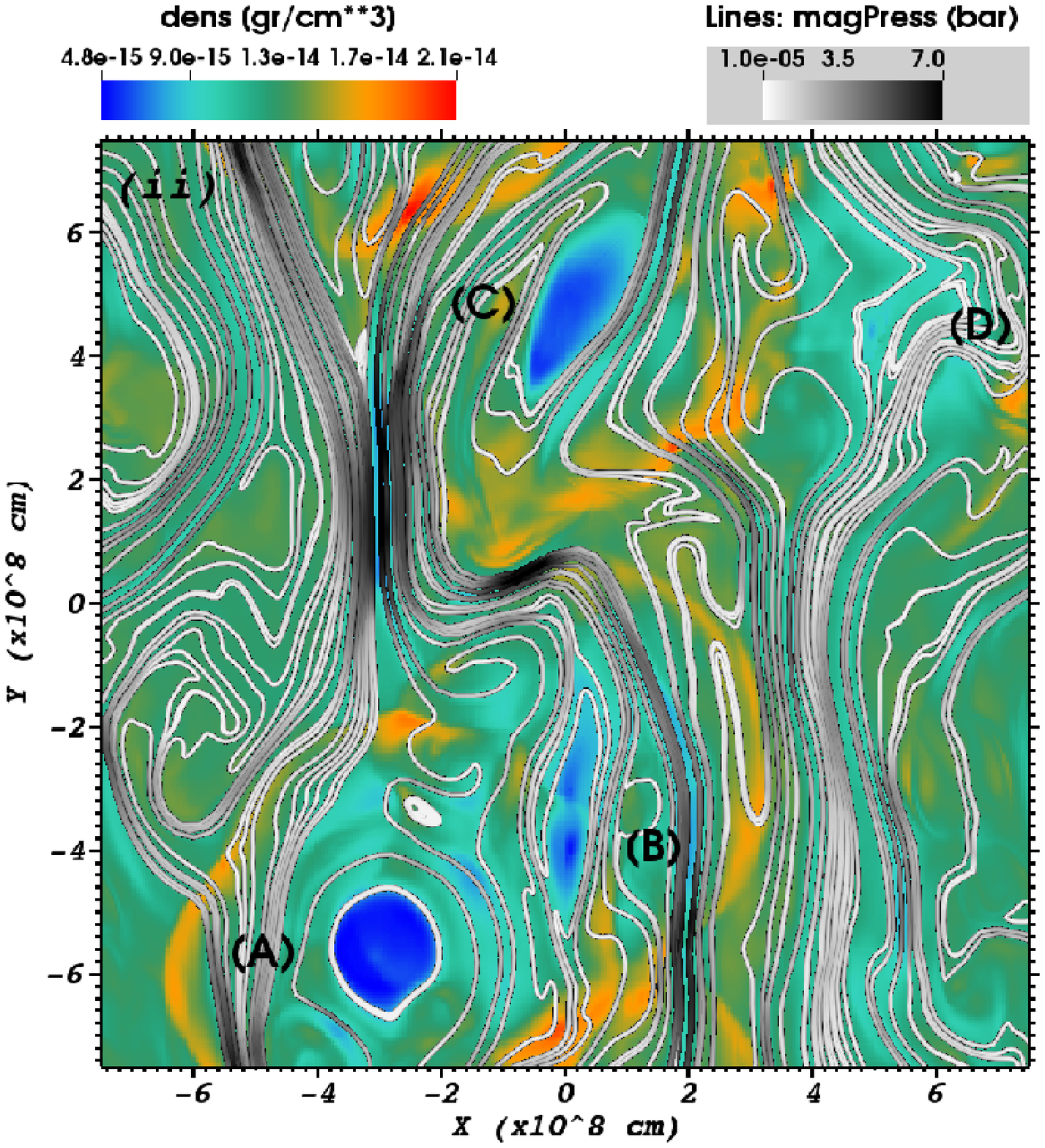}
\includegraphics[width=0.4\textwidth]{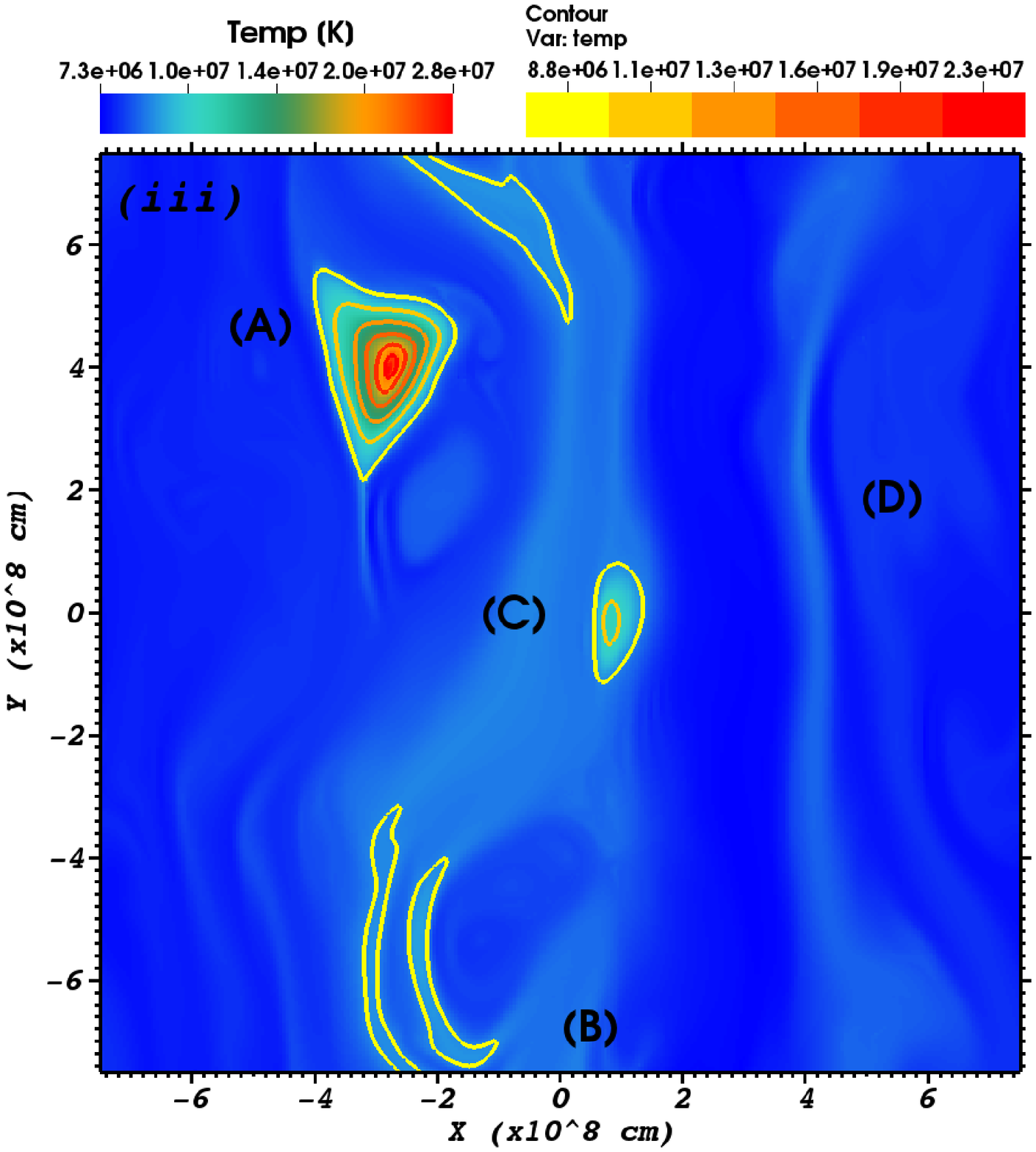}
\includegraphics[width=0.4\textwidth]{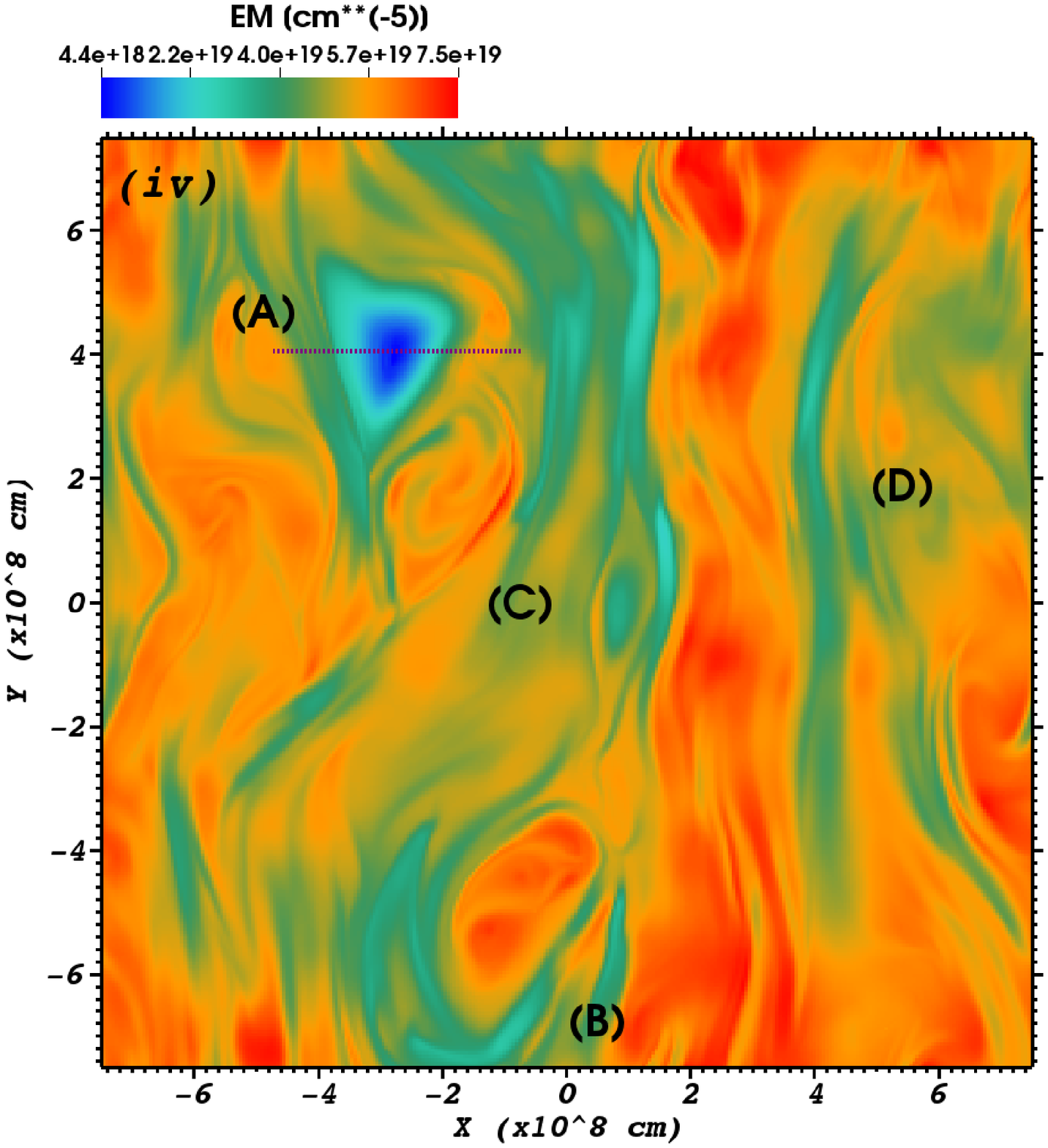}
\caption{(i) Temperature at $t=312~$s, $(A)$, $(B)$, $(C)$ and $(D)$ are cavities triggered at $t=305.9~$s. (ii) Density at $t=312$s. Same as before but with magnetic field lines superimposed. (iii) Same as (i) one minute later, at $t=372~$s. Associated with this figure we show an online animated figure with partially superimposed magnetic field lines. (iv) EM, also at $t=372\,$s (the dotted line corresponds to EM cut line shown in Fig. (\ref{f.em_corte}) at $t=372~$s).  Only cavity $(A)$ has an EM contrast of $\gtrsim 4$, $(B)$ scribbled, $(C)$ smoothed out and $(D)$ faded away.  (Parallel magnetic field case).} 
\label{f.sad}
\end{center}
\end{figure}

The instantaneous pressure pulses are applied at time $t=305.9\,$s. Figure \ref{f.sad} shows in (i) the temperature pattern and in (ii) the density with overlapped magnetic field lines, at an early time $t=312\,$s. The resulting pulse patterns are labelled by $(A)$, $(B)$, $(C)$ and $(D)$. Figure \ref{f.sad}(iii-iv) shows the same evolved patterns a minute later, at $t=372\,$s, for the temperature with temperature contours superimposed (iii) and the EM (iv). Associated to Figure \ref{f.sad}(iii) we include an online animated figure for the temperature pattern during the  fan stage. The EM contrast of the SAD candidates will fade away fast and before the desired time if during their evolution they are not sufficiently isolated by the magnetic field. In other words, to prevail for a significant time-span they need to be included inside an island. This is not the case for features $(B)$, $(C)$ and $(D)$ that can not be identified as SADs, according to our requirements. Particularly, only vestiges of feature $(B)$ remain and a slight hot subdense lane of feature $(D)$ can be appreciated at $x\approx 4\times 10^8~$cm. Seeing the contour curves enclosing the SAD, note that the temperature gradient is perpendicular to the magnetic field lines, thermally isolating the SAD structure. The different resulting dynamic evolutions of the bursty reconnections, $(A)-(D)$, necessarily resides in the degree of thermal isolation that the  magnetic topology provides during  their time evolution. 

\begin{figure}
\centering
\includegraphics[width=0.4\textwidth]{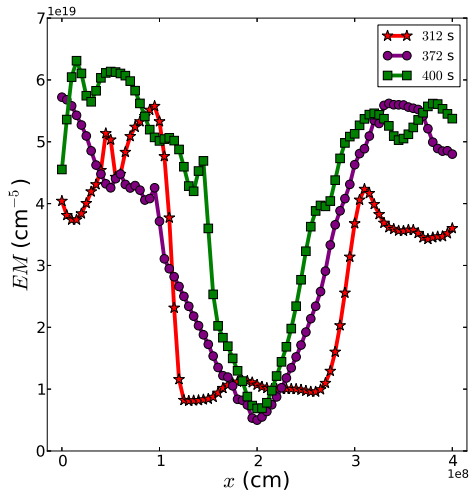}
\caption{Cut of EM contrast of cavity $(B)$ at times $t=312~$s, $t=372~$s (dotted line in Fig. (\ref{f.sad})(iv)) and $t=398~$s. } 
\label{f.em_corte}
\end{figure}

Figure \ref{f.em_corte} shows an EM cut over cavity $(A)$ (dotted line in Fig. \ref{f.sad}(iv)) at times $t=312~$s, $372~$s and $398~$s, respectively. The totally enclosed cavity $(A)$ is maintained along time with an EM contrast enough to allow a SAD observation (see figure 4(b) in \citet{2012ApJ...747L..40S}). 

As mentioned, we implemented an external mechanism to develop the turbulent state  based on an observational established turbulent fan description \citep{2013ApJ...766...39M, 2016ApJ...819...56S,2012ApJ...747L..40S}.  While not conclusive, SAD detections  mostly occur during the decay phase of the flaring activity. Thus, the  artificial forcing turn off could be in line with the flaring decay leading the system to a decaying turbulent regime with magnetic field lines straightening due to the action of the magnetic tension. At the same time, in the framework of our scenario, downward velocity shears are necessary to explain the motion of the observed low-density regions that can be associated with the action of jets and outflows coming from upper CS structures \citep{2012ApJ...754...13S}. Shears usually result in  KH instabilities that  favour reconnection and consequently the production of isolated islands within the fan \citep{2008JGRA..113.9204N}. While shears are a model condition to give account of SADs velocities we are now interested in the generation and survival of island topologies by KH instabilities as it seems to be an important factor for SAD observability.
 
\citet{2008JGRA..113.9204N} studied the coupling between MHD scales with KH vortices and spontaneous magnetic reconnection. They started from a homogeneous $2$D two-fluid plasma as initial condition and found  two  reconnection types that occur associated with KH vortices. One type is related with in-plane perpendicular magnetic fields across the velocity shear layer. The other is driven when the velocity shear is strong enough to produce highly rolled-up vortices and highly stretched inner field lines leading to the formation of reconnected islands.  They also found that reconnection is triggered even for small values of the Alfv\'en Mach number ($M_A$) of the shear, i.e., $2<M_A<5$. A difference with these authors is that our initial condition is an already turbulent inhomogeneous one fluid plasma state with a shear $M_A\simeq 1.3$. For simplicity we have not included a physical resistivity term in the simulations but we assume that the numerical diffusion that artificially  reconnects close field lines provides the necessary reconnection (it actually over-estimates the reconnection as we are not able to reach the diffusion scale-lengths). To qualitatively study the coupling  between KH and reconnection leading to an enhance of island production we compare the stationary turbulent state (at the beginning of the fan stage, at $t=303.8~$s), with the immediate state after the shear. 
  
\begin{figure} 
\begin{center}
\includegraphics[width=0.32\textwidth]{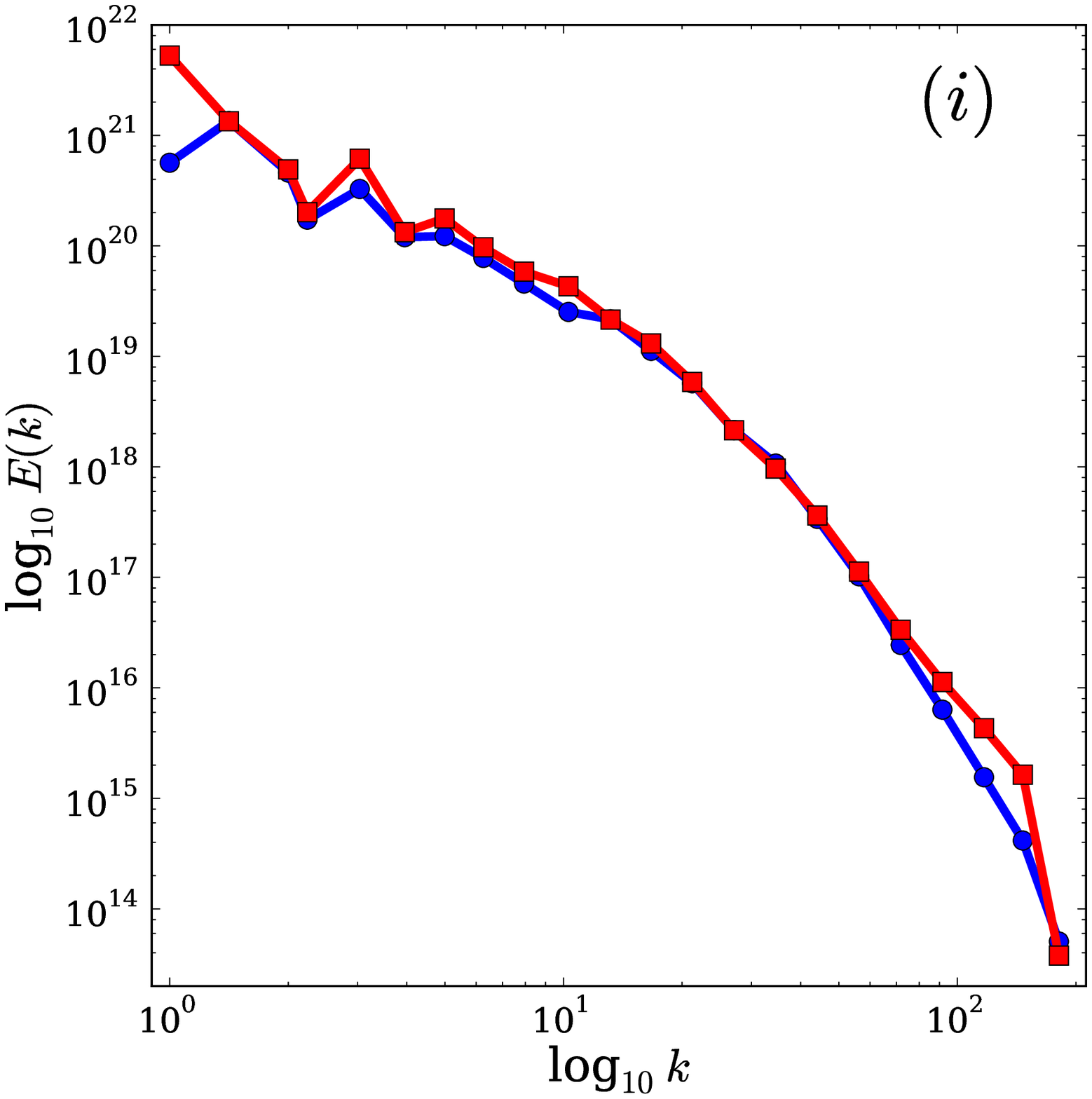}
\includegraphics[width=0.32\textwidth]{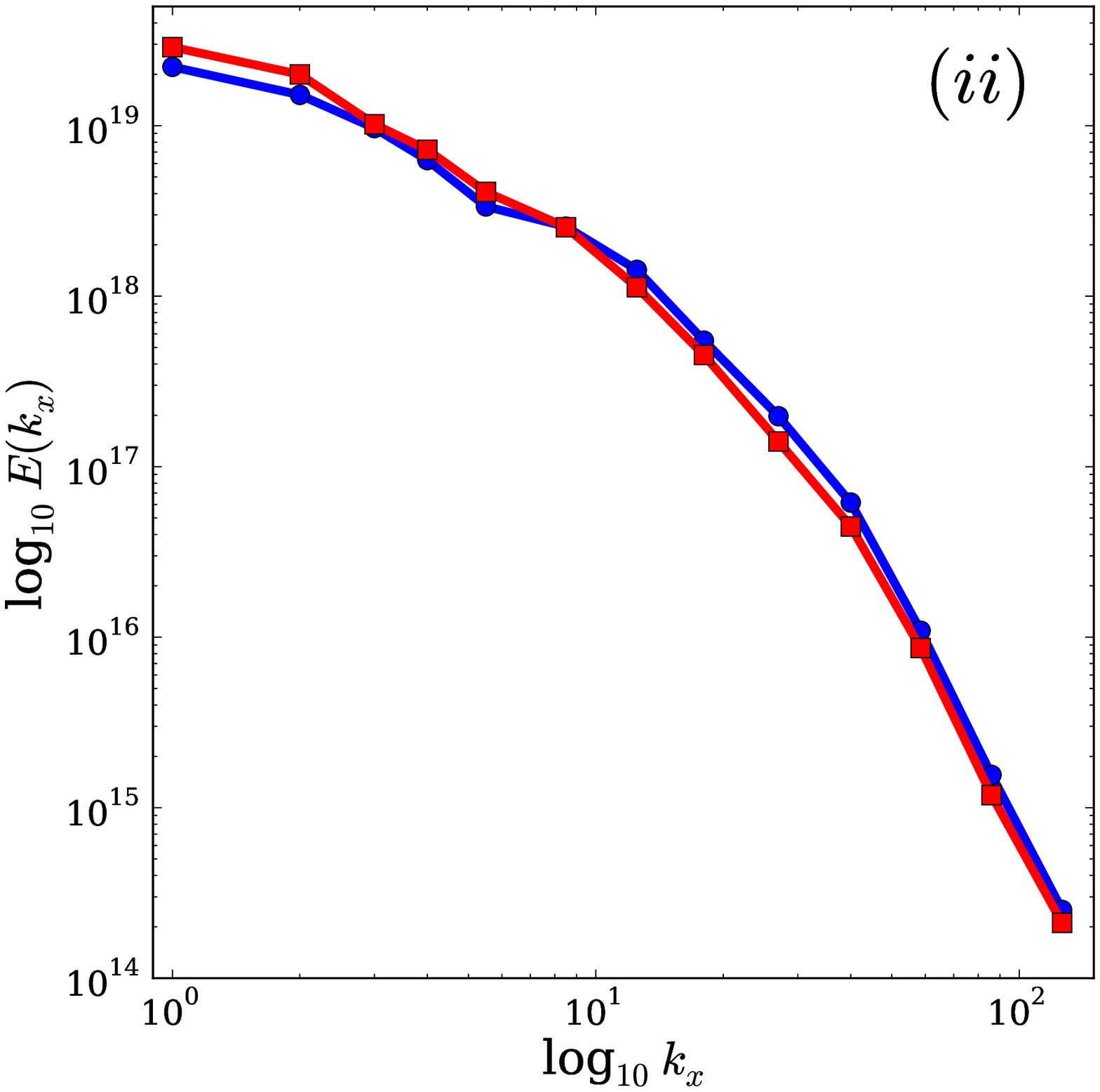}
\includegraphics[width=0.32\textwidth]{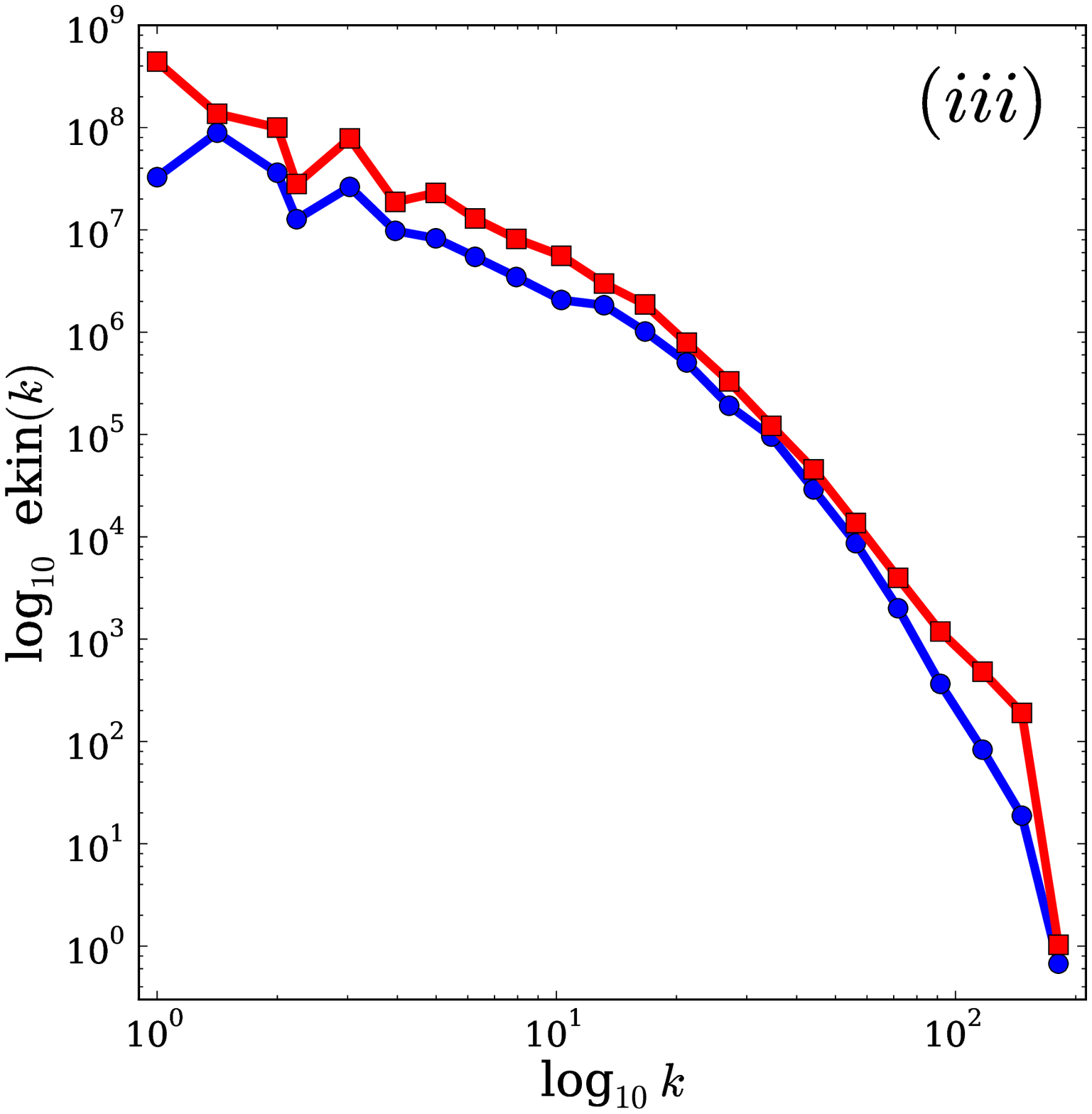}
\caption{(i) Power spectra of the velocity $E (k)$ as a function of $k$. (ii) Velocity $E (k_x)$ as a function of $k_x$. (iii) Kinetic energy $ekin (k)$ as a function of $k$. For all cases the red line represents the fan stage at $t=303.0~$s just before the shear is applied, and the blue one just after it, at $t=304.0~$s. } 
\label{f.KH}
\end{center}
\end{figure}

Figure \ref{f.KH} displays the power spectrum (in Fourier space) of the velocity $E(k)$ (i and ii) and the kinetic energy (iii) just before ($t=303.0~$s) and after ($t=304.0~$s) the shear is applied (Eq. \ref{10}). In this analysis anisotropic TC is not considered. Figure \ref{f.KH}(i) shows the velocity power spectrum as a function of $|k|$. The instantaneous shear implies a sudden increase of the curve for the smallest value of $|k|$. Still, the action of the stress tensors yield a contribution that propagates rapidly to the diffusion scales, i.e., at large values of $|k|$. One and a half minutes later (at the end of the fan stage, $t=400~$s), a typical time of a SAD life, these curves are virtually the same (not showed in the figure). The shear results in an anisotropic contribution to the velocity along the $y$-direction, we accordingly find that the contribution of the velocity strength along $k_x$ is almost the same for all times of interest (Fig. \ref{f.KH}(ii)) meaning that the differences between the curves of  Fig. \ref{f.KH}(i) are due to this anisotropic and coherent velocity strength. The Fourier transform as a function of $k_y$ (not shown) is almost the same as Fig. \ref{f.KH}(i).  Figure \ref{f.KH}(iii) shows the kinetic energy power spectrum as a function of $|k|$, the differences between the energy contribution of the modes in the cases with and without shear are more pronounced, but still rather small. We speculate that the action of the stress tensor (when the shear is applied) leads to an increase of mode contributions at all scales, but in particular near the diffusion region increasing the reconnection rate, this is in agreement with \citet{2008JGRA..113.9204N} who reported an enhance of islands production when shear was present.  

\begin{figure} 
\begin{center}
\includegraphics[width=0.4\textwidth]{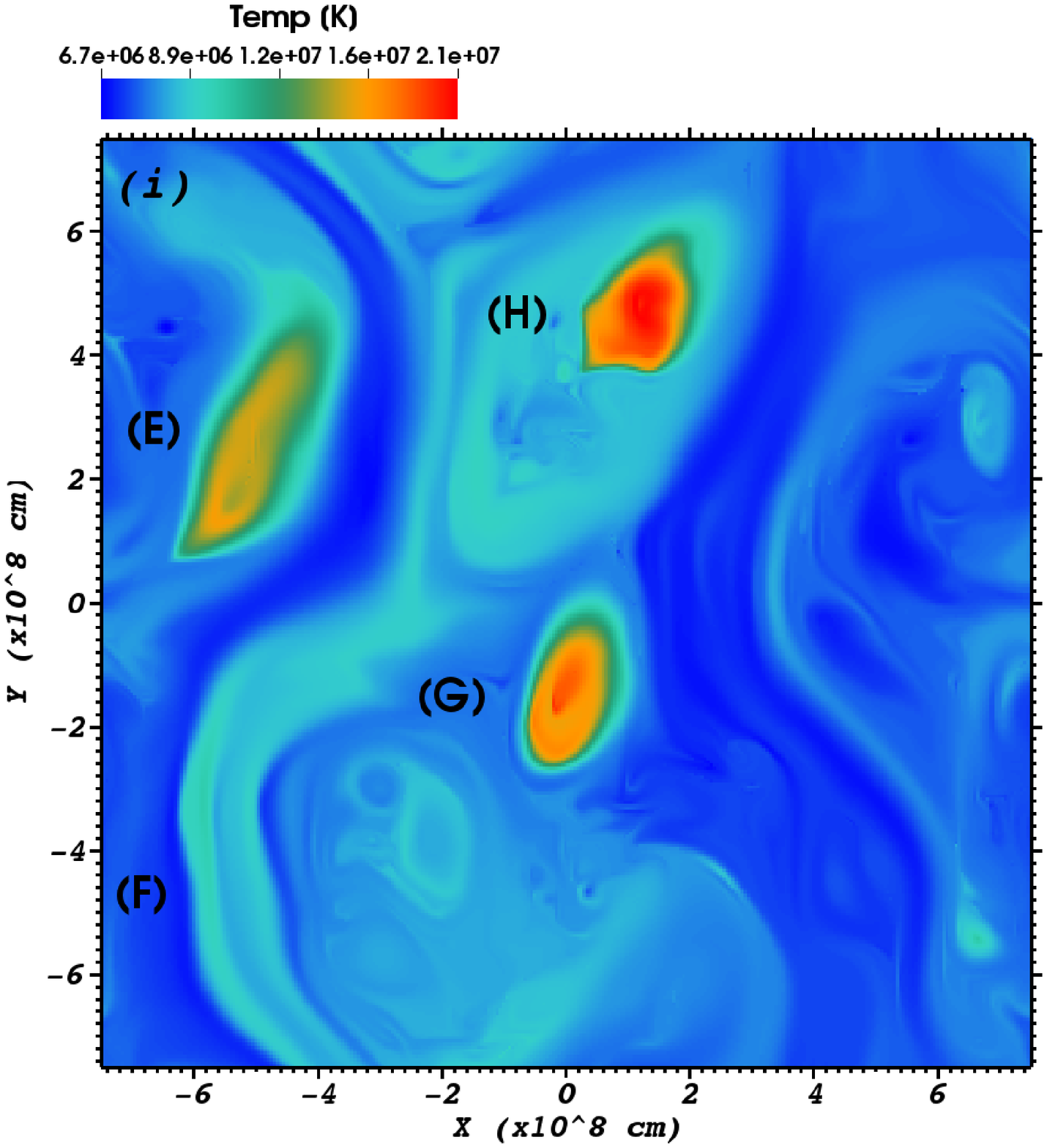}
\includegraphics[width=0.4\textwidth]{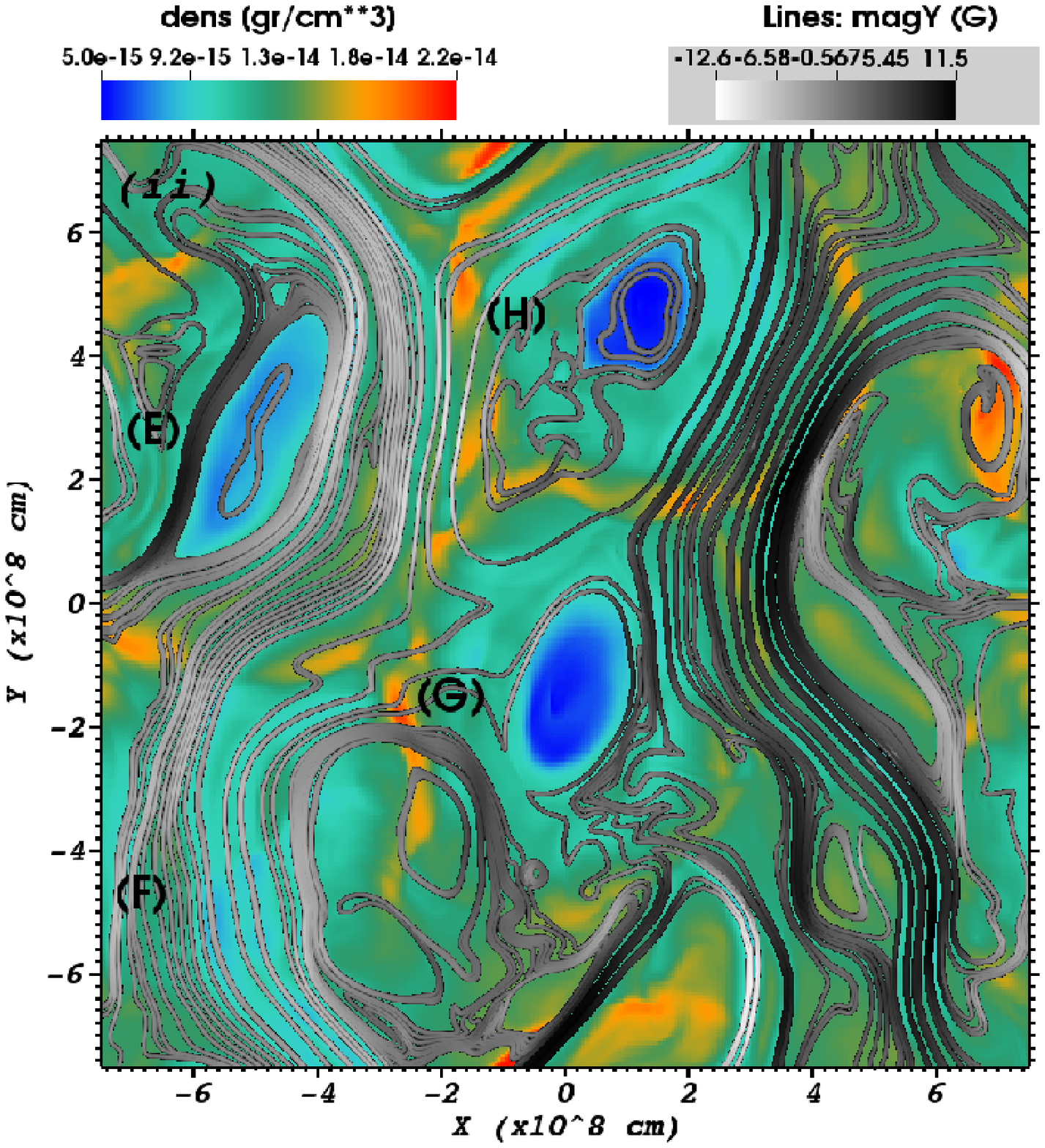}
\includegraphics[width=0.4\textwidth]{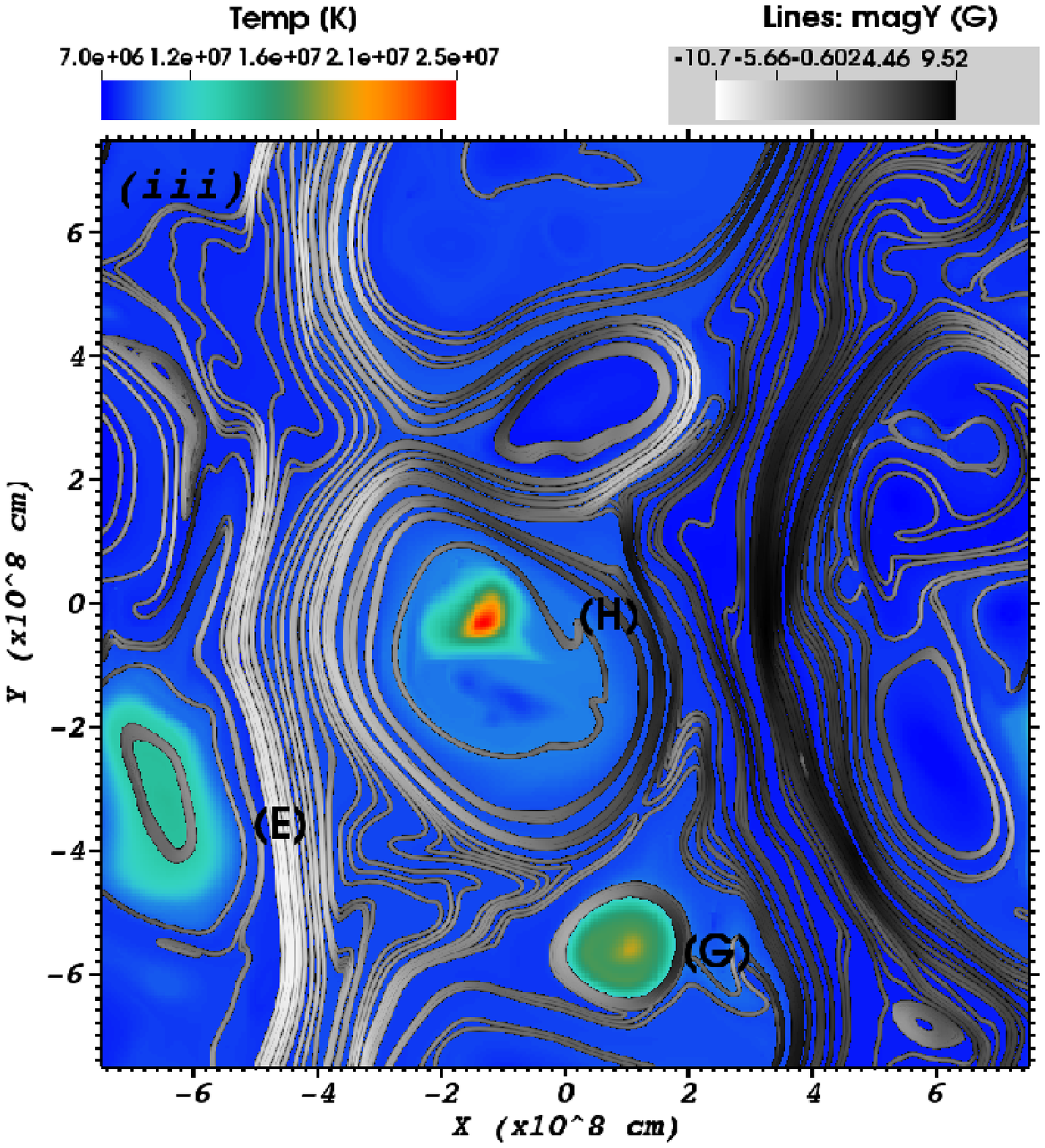}
\includegraphics[width=0.4\textwidth]{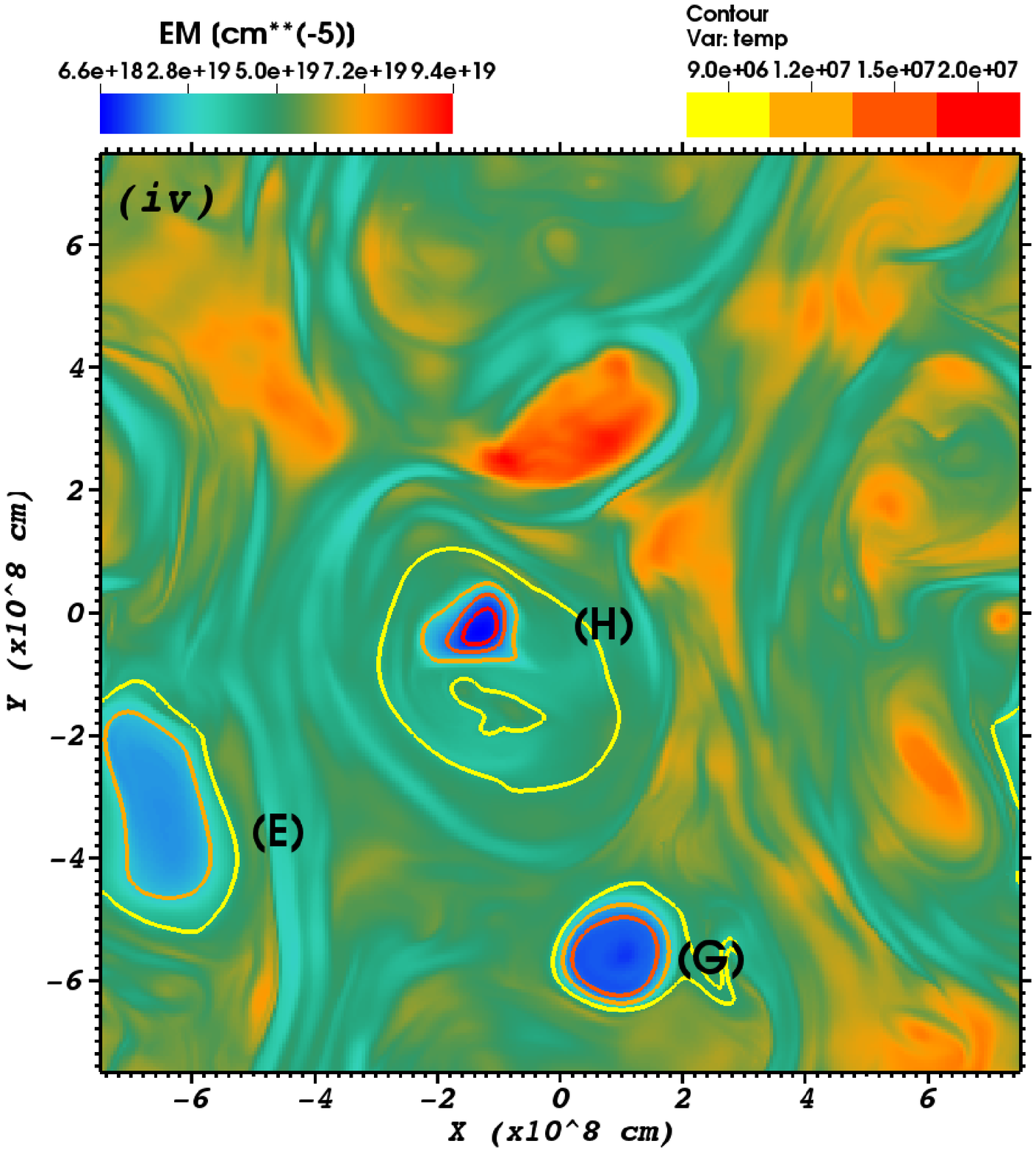}
\caption{(i) Temperature at $t=312~$s, $(E)$, $(F)$, $(G)$ and $(H)$ labelled cavities triggered at $t=305.9~$s. (ii) Density at $t=312~$s. Same as before but with magnetic field lines superimposed. (iii) Same as (i) one minute later, at $t=372~$s. (iv) EM, also at $t=372~$s. Cavities $(G)$ and $(H)$ have an EM contrast of $\gtrsim 4$, $(F)$ faded away and $(E)$ smoothed out. (Anti-parallel magnetic field case).}
\label{f.antip}
\end{center}
\end{figure}

If we consider a $3D$ magnetic configuration scheme (see figure 1 of \citetalias{2015ApJ...807....6C}), the setup described by Table (\ref{table.uno}) and whose results are shown in Fig. (\ref{f.sad}), will represent the face-on view. To also consider the edge-on view, we perform another simulation where the initial magnetic field instead of being homogeneous is the CS given by 
\begin{equation}
B_{y}(x) = \left\{
\begin{array}{l l}
\ \ B_{0}  & \quad \mbox{if $ x < 0 $,} \\
-B_{0}  & \quad \mbox{if $ x \ge 0$, }
\end{array} \right.  \label{11}
\end{equation}
with $B_{0}=3.0~$G. The preparation of the fan is performed  with the two stages as in the previous case, but in this case the shear (Eq. (\ref{10})) coinciding with the CS location. The instantaneous pressure pulses are again applied at time $t=305.9~$s. Figure \ref{f.antip} shows the temperature (i) and the density with overlapped magnetic field lines (ii) at an early time $t=312~$s. The resulting pulse patterns are labelled by $(E)$, $(F)$, $(G)$ and $(H)$. Figure (\ref{f.antip})(iii-iv) shows the same evolved patterns a minute later, at $t=372~$s, for the temperature with magnetic field lines superimposed (iii) and the EM with temperature contours (iv). Same as before, the EM contrast of the SAD candidates will fade away fast and before the desired time if during their evolution they are not sufficiently isolated by the magnetic field. In fact the feature $(F)$ triggered in a location where the magnetic field lines are almost straight is not appreciated even at an early time (Fig. \ref{f.antip}(i-ii)).  The cavities that survive and can be identified as SADs are located mainly in the neighborhood of the CS\footnote{This is noted clearer when the turbulence stage is carried on using larger wave-numbers comparing with those chosen here. See also \citetalias{2015ApJ...807....6C}.} where the magnetic field is rolled-up forming closed vortices and islands.

The observations have generally shown that SADs are distributed along the whole fan and not necessarily concentrated in the CS plane -which may indicate that shears are not always coincident with the CS plane-, and that SADs are preferentially detected in face-on views rather than in edge-on ones. Within the current $2D$ scheme, we conclude that the setup representing the face-on view (Fig. \ref{f.sad}) and the edge-on one (Fig. \ref{f.antip}) are both able to produce detectable SADs satisfying the above requirements, where the face-on case seems to be more general and less restrictive than the edge-one one. However, in the $3D$ scheme the electronic number density also depends on the $z$-direction. Therefore, as the CS is much longer (along the arcade axis) than thicker (perpendicular to the arcade axis), to get an observable SAD the column plasma (or fan) and the subdense cavity characteristic longitudes must be approximately of the same order (\citetalias{2015ApJ...807....6C}). Thus, the EM contrast requirement ($\gtrsim 4$) is much likely to be satisfied in the face-on view than in the edge-on one.  
%

\section{Conclusions} \label{s.conclusions}

Supra-arcade downflows (SADs) are known to be subdense dark moving trails in turbulent fan regions above flaring arcades.
These dynamic features, generally coming from the upper and cooler background corona, move twisting between fan interstices of inhomogeneous density and magnetic field. Several scenarios have been proposed to account for the nature and dynamical origin of SADs. In previous works we proposed a scenario where SADs are cavities created by the nonlinear waves disturbing the density, and lasting for times comparable with observations. They are triggered by bursty localized reconnection events occurring either in the upper background corona or/and in the nearer fan region. The different proposed scenarios have generally eluded the systematic consideration of thermal conduction effects of these structures moving in  hot and large temperature gradient fans $(T\gtrsim 7)~$MK, which would tend to diffuse the SADs on time-scales shorter than their observed life-time.  Here we analysed different plasma conditions, considering anisotropic thermal conduction, that make possible the survival of SADs avoiding the usually strong thermal damping. We consider a SAD as subdense voided cavities with a radius of $\sim 1~$Mm  (almost the smallest values observed),  that can maintain emission measure contrast values $\gtrsim 4$ and that prevail for at least one minute in a hot $2$D fan medium. We found that triggered subdense cavities moving in a turbulent vortical magnetized fan can be identified as SAD features if in their downward motion they become enveloped by the magnetic field in such a way that heat conduction is inhibited. We also found that an increase of magnetic island production, where SADs are  completely thermally isolated, reinforces the life-time duration of SADs. Islands are expected  due to shears produced by upper observational current sheet outflows and jets that lead to  Kelvin-Helmoholtz  instabilities allowing the reconnection of close enough magnetic field lines. The large fan values of $\beta$ found are in agreement with observations, also reinforces the turbulence development, i.e., the larger the $\beta$ values, the gas pressure more efficiently will bend the magnetic field lines. This allows the plasma envelope into highly rolled-up vortices and totally isolated islands preventing the action of the thermal conduction diffusion.

\section*{Acknowledgements}
This work has been supported by CONICET (Argentina) through a PhD grant. A. Esquivel acknowledges support from CONACYT (Mexico) grant 167611, and DGAPA-PAPIIT (UNAM) grants IN 109715 and IG-RG 100516. The software used in this work was in part developed by the ASC/Alliance Center for Astrophysical Thermonuclear Flashes at the University of Chicago. Also we would like to thank to the VisIt team.


\bibliography{zurbriggen_e} 

\begin{thebibliography}{}
\expandafter\ifx\csname natexlab\endcsname\relax\def\natexlab#1{#1}\fi

\bibitem[{{Asai} {et~al.}(2004){Asai}, {Yokoyama}, {Shimojo}, \&
  {Shibata}}]{2004ApJ...605L..77A}
{Asai}, A., {Yokoyama}, T., {Shimojo}, M., \& {Shibata}, K. 2004, \apjl, 605,
  L77

\bibitem[{{Aschwanden}(2005)}]{2005psci.book.....A}
{Aschwanden}, M.~J. 2005, {``Physics of the Solar Corona. An Introduction with
  Problems and Solutions''} (Praxis Publishing Ltd., 2nd ed.|2005)

\bibitem[{{Bemporad}(2008)}]{2008ApJ...689..572B}
{Bemporad}, A. 2008, \apj, 689, 572

\bibitem[{{C{\'e}cere} {et~al.}(2012){C{\'e}cere}, {Schneiter}, {Costa},
  {Elaskar}, \& {Maglione}}]{2012ApJ...759...79C}
{C{\'e}cere}, M., {Schneiter}, M., {Costa}, A., {Elaskar}, S., \& {Maglione},
  S. 2012, \apj, 759, 79

\bibitem[{{C{\'e}cere} {et~al.}(2015){C{\'e}cere}, {Zurbriggen}, {Costa}, \&
  {Schneiter}}]{2015ApJ...807....6C}
{C{\'e}cere}, M., {Zurbriggen}, E., {Costa}, A., \& {Schneiter}, M. 2015, \apj,
  807, 6

\bibitem[{{Chen} \& {Lykoudis}(1972)}]{1972SoPh...25..380C}
{Chen}, C.-J., \& {Lykoudis}, P.~S. 1972, \solphys, 25, 380

\bibitem[{{Costa} {et~al.}(2009){Costa}, {Elaskar}, {Fern{\'a}ndez}, \&
  {Mart{\'{\i}}nez}}]{2009MNRAS.400L..85C}
{Costa}, A., {Elaskar}, S., {Fern{\'a}ndez}, C.~A., \& {Mart{\'{\i}}nez}, G.
  2009, \mnras, 400, L85

\bibitem[{{Cowie} \& {McKee}(1977)}]{1977ApJ...211..135C}
{Cowie}, L.~L., \& {McKee}, C.~F. 1977, \apj, 211, 135

\bibitem[{{Eswaran} \& {Pope}(1988)}]{1988CF.....16..257E}
{Eswaran}, V., \& {Pope}, S.~B. 1988, Computers and Fluids, 16, 257

\bibitem[{{Federrath} {et~al.}(2010){Federrath}, {Roman-Duval}, {Klessen},
  {Schmidt}, \& {Mac Low}}]{2010A&A...512A..81F}
{Federrath}, C., {Roman-Duval}, J., {Klessen}, R.~S., {Schmidt}, W., \& {Mac
  Low}, M.-M. 2010, \aap, 512, A81

\bibitem[{{Fryxell} {et~al.}(2000){Fryxell}, {Olson}, {Ricker}, {Timmes},
  {Zingale}, {Lamb}, {MacNeice}, {Rosner}, {Truran}, \&
  {Tufo}}]{2000ApJS..131..273F}
{Fryxell}, B., {Olson}, K., {Ricker}, P., {et~al.} 2000, \apjs, 131, 273

\bibitem[{{Goldreich} \& {Sridhar}(1995)}]{1995ApJ...438..763G}
{Goldreich}, P., \& {Sridhar}, S. 1995, \apj, 438, 763

\bibitem[{{Guo} {et~al.}(2014){Guo}, {Huang}, {Bhattacharjee}, \&
  {Innes}}]{2014ApJ...796L..29G}
{Guo}, L.-J., {Huang}, Y.-M., {Bhattacharjee}, A., \& {Innes}, D.~E. 2014,
  \apjl, 796, L29

\bibitem[{{Hanneman} \& {Reeves}(2014)}]{2014ApJ...786...95H}
{Hanneman}, W.~J., \& {Reeves}, K.~K. 2014, \apj, 786, 95

\bibitem[{{Hensler} \& {Vieser}(2002)}]{2002Ap&SS.281..275H}
{Hensler}, G., \& {Vieser}, W. 2002, \apss, 281, 275

\bibitem[{{Innes} {et~al.}(2014){Innes}, {Guo}, {Bhattacharjee}, {Huang}, \&
  {Schmit}}]{2014ApJ...796...27I}
{Innes}, D.~E., {Guo}, L.-J., {Bhattacharjee}, A., {Huang}, Y.-M., \& {Schmit},
  D. 2014, \apj, 796, 27

\bibitem[{{Lazarian} \& {Vishniac}(1999)}]{1999ApJ...517..700L}
{Lazarian}, A., \& {Vishniac}, E.~T. 1999, \apj, 517, 700

\bibitem[{{Lee} {et~al.}(2009){Lee}, {Deane}, \&
  {Federrath}}]{2009ASPC..406..243L}
{Lee}, D., {Deane}, A.~E., \& {Federrath}, C. 2009, in Astronomical Society of
  the Pacific Conference Series, Vol. 406, Numerical Modeling of Space Plasma
  Flows: ASTRONUM-2008, ed. N.~V. {Pogorelov}, E.~{Audit}, P.~{Colella}, \&
  G.~P. {Zank}, 243

\bibitem[{{Linton} {et~al.}(2009){Linton}, {Devore}, \&
  {Longcope}}]{2009EP&S...61..573L}
{Linton}, M.~G., {Devore}, C.~R., \& {Longcope}, D.~W. 2009, Earth, Planets,
  and Space, 61, 573

\bibitem[{{Liu} {et~al.}(2013){Liu}, {Chen}, \&
  {Petrosian}}]{2013ApJ...767..168L}
{Liu}, W., {Chen}, Q., \& {Petrosian}, V. 2013, \apj, 767, 168

\bibitem[{{McKenzie}(2013)}]{2013ApJ...766...39M}
{McKenzie}, D.~E. 2013, \apj, 766, 39

\bibitem[{{McKenzie} \& {Hudson}(1999)}]{1999ApJ...519L..93M}
{McKenzie}, D.~E., \& {Hudson}, H.~S. 1999, \apjl, 519, L93

\bibitem[{{McKenzie} \& {Savage}(2009)}]{2009ApJ...697.1569M}
{McKenzie}, D.~E., \& {Savage}, S.~L. 2009, \apj, 697, 1569

\bibitem[{{Nakamura} {et~al.}(2008){Nakamura}, {Fujimoto}, \&
  {Otto}}]{2008JGRA..113.9204N}
{Nakamura}, T.~K.~M., {Fujimoto}, M., \& {Otto}, A. 2008, Journal of
  Geophysical Research (Space Physics), 113, A09204

\bibitem[{{Pagano} {et~al.}(2007){Pagano}, {Reale}, {Orlando}, \&
  {Peres}}]{2007A&A...464..753P}
{Pagano}, P., {Reale}, F., {Orlando}, S., \& {Peres}, G. 2007, \aap, 464, 753

\bibitem[{{Prialnik}(2000)}]{2000itss.book.....P}
{Prialnik}, D. 2000, {``An Introduction to the Theory of Stellar Structure and
  Evolution''} (Cambridge: University Press, 2000)

\bibitem[{{Savage} {et~al.}(2012{\natexlab{a}}){Savage}, {Holman}, {Reeves},
  {Seaton}, {McKenzie}, \& {Su}}]{2012ApJ...754...13S}
{Savage}, S.~L., {Holman}, G., {Reeves}, K.~K., {et~al.} 2012{\natexlab{a}},
  \apj, 754, 13

\bibitem[{{Savage} {et~al.}(2012{\natexlab{b}}){Savage}, {McKenzie}, \&
  {Reeves}}]{2012ApJ...747L..40S}
{Savage}, S.~L., {McKenzie}, D.~E., \& {Reeves}, K.~K. 2012{\natexlab{b}},
  \apjl, 747, L40

\bibitem[{{Scott} {et~al.}(2016){Scott}, {McKenzie}, \&
  {Longcope}}]{2016ApJ...819...56S}
{Scott}, R.~B., {McKenzie}, D.~E., \& {Longcope}, D.~W. 2016, \apj, 819, 56

\bibitem[{{Seaton} \& {Forbes}(2009)}]{2009ApJ...701..348S}
{Seaton}, D.~B., \& {Forbes}, T.~G. 2009, \apj, 701, 348

\bibitem[{{Spitzer}(1962)}]{1962pfig.book.....S}
{Spitzer}, L. 1962, {``Physics of Fully Ionized Gases''} (New York:
  Interscience, 2nd ed.|1962)

\bibitem[{{Yokoyama} \& {Shibata}(2001)}]{2001ApJ...549.1160Y}
{Yokoyama}, T., \& {Shibata}, K. 2001, \apj, 549, 1160

\end{thebibliography}

\end{document}